\documentclass[longauth]{aa}  
\usepackage{hyperref}

\usepackage{graphicx}
\usepackage{txfonts}
\usepackage{xspace}
\usepackage{booktabs}
\usepackage{makecell}
\usepackage{multirow}
\usepackage{threeparttable}
\usepackage{pifont}
\usepackage{newtxmath}
\usepackage[usenames,dvipsnames]{color}
\usepackage[modulo, switch]{lineno}
\usepackage{enumitem}
\setlist{nolistsep}

\newcommand{\kms}{km\,s$^{-1}$\xspace}
\newcommand{\zerr}{\texttt{ZERR}\xspace}

\usepackage{natbib}
\bibpunct{(}{)}{;}{a}{}{,} 
\begin{document} 

\title{Peering down the barrel with DESI DR2: 10 000+ inflows at $z < 0.6$ reveal how galaxies accrete cold gas}

\author{S.~Weng \inst{1}\fnmsep\thanks{Corresponding author: \email{simon.weng@lam.fr}} \and
          A.~Saintonge \inst{2,3} \and
          M.~Pieri \inst{1} \and
          J.~Moustakas \inst{4} \and
          H.~Zou \inst{5} \and
          D.~Mu\~noz Santos \inst{1} \and
          J.~Yu \inst{6} \and
          J.~Aguilar \inst{7} \and
          S.~Ahlen \inst{8} \and
          D.~Bianchi \inst{9,10} \and
          D.~Brooks \inst{2} \and
          T.~Claybaugh \inst{7} \and
          A.~Cuceu \inst{7} \and
          A.~de la Macorra \inst{11} \and
          P.~Doel \inst{2} \and
          A.~Font-Ribera \inst{12,13} \and
          J.~E.~Forero-Romero \inst{14,15} \and
          E.~Gaztañaga \inst{16,17,18} \and
          S.~{Gontcho A Gontcho} \inst{19} \and
          G.~Gutierrez \inst{20} \and
          C.~Hahn \inst{21} \and
          S.~He \inst{22} \and
          K.~Honscheid \inst{23,24,25} \and
          T.~Hu \inst{1} \and
          R.~Joyce \inst{26} \and
          R.~Kehoe \inst{27} \and
          M.~Landriau \inst{7} \and
          L.~Le~Guillou \inst{28} \and
          A.~Meisner \inst{26} \and
          R.~Miquel \inst{12,13} \and
          S.~Nadathur \inst{17} \and
          J.~ A.~Newman \inst{29} \and
          W.~J.~Percival \inst{30,31,32} \and
          I.~P\'erez-R\`afols \inst{33} \and
          G.~Rossi \inst{34} \and
          E.~Sanchez \inst{35} \and
          D.~Schlegel \inst{7} \and
          H.~Seo \inst{36} \and
          J.~Silber \inst{7} \and
          D.~Sprayberry \inst{26} \and
          G.~Tarl\'{e} \inst{37} \and
          B.~A.~Weaver \inst{26}
         }
\institute{Aix Marseille Univ, CNRS, CNES, LAM, Marseille, France
    \and
    Department of Physics \& Astronomy, University College London, Gower Street, London, WC1E 6BT, UK
    \and
    Max-Planck-Institut f\"{u}r Radioastronomie, Auf dem H\"{u}gel 69, 53121 Bonn, Germany
    \and
    Department of Physics and Astronomy, Siena University, 515 Loudon Road, Loudonville, NY 12211, USA
    \and
    National Astronomical Observatories, Chinese Academy of Sciences, A20 Datun Road, Chaoyang District, Beijing, 100101, P.~R.~China
    \and
    Kavli Institute for the Physics and Mathematics of the Universe (WPI), The University of Tokyo Institutes for Advanced Study (UTIAS), The University of Tokyo, Chiba 277-8583, Japan
    \and
    Lawrence Berkeley National Laboratory, 1 Cyclotron Road, Berkeley, CA 94720, USA
    \and
    Department of Physics, Boston University, 590 Commonwealth Avenue, Boston, MA 02215 USA
    \and
    Dipartimento di Fisica ``Aldo Pontremoli'', Universit\`a degli Studi di Milano, Via Celoria 16, I-20133 Milano, Italy
    \and
    INAF-Osservatorio Astronomico di Brera, Via Brera 28, 20122 Milano, Italy
    \and
    Instituto de F\'{\i}sica, Universidad Nacional Aut\'{o}noma de M\'{e}xico,  Circuito de la Investigaci\'{o}n Cient\'{\i}fica, Ciudad Universitaria, Cd. de M\'{e}xico  C.~P.~04510,  M\'{e}xico
    \and
    Instituci\'{o} Catalana de Recerca i Estudis Avan\c{c}ats, Passeig de Llu\'{\i}s Companys, 23, 08010 Barcelona, Spain
    \and
    Institut de F\'{i}sica d’Altes Energies (IFAE), The Barcelona Institute of Science and Technology, Edifici Cn, Campus UAB, 08193, Bellaterra (Barcelona), Spain
    \and
    Departamento de F\'isica, Universidad de los Andes, Cra. 1 No. 18A-10, Edificio Ip, CP 111711, Bogot\'a, Colombia
    \and
    Observatorio Astron\'omico, Universidad de los Andes, Cra. 1 No. 18A-10, Edificio H, CP 111711 Bogot\'a, Colombia
    \and
    Institut d'Estudis Espacials de Catalunya (IEEC), c/ Esteve Terradas 1, Edifici RDIT, Campus PMT-UPC, 08860 Castelldefels, Spain
    \and
    Institute of Cosmology and Gravitation, University of Portsmouth, Dennis Sciama Building, Portsmouth, PO1 3FX, UK
    \and
    Institute of Space Sciences, ICE-CSIC, Campus UAB, Carrer de Can Magrans s/n, 08913 Bellaterra, Barcelona, Spain
    \and
    University of Virginia, Department of Astronomy, Charlottesville, VA 22904, USA
    \and
    Fermi National Accelerator Laboratory, PO Box 500, Batavia, IL 60510, USA
    \and
    Department of Astronomy, University of Texas at Austin, 2515 Speedway, TX 78712, USA
    \and
    Institute of Physics, Laboratory of Astrophysics, \'{E}cole Polytechnique F\'{e}d\'{e}rale de Lausanne (EPFL), Observatoire de Sauverny, Chemin Pegasi 51, CH-1290 Versoix, Switzerland
    \and
    Center for Cosmology and AstroParticle Physics, The Ohio State University, 191 West Woodruff Avenue, Columbus, OH 43210, USA
    \and
    Department of Physics, The Ohio State University, 191 West Woodruff Avenue, Columbus, OH 43210, USA
    \and
    The Ohio State University, Columbus, 43210 OH, USA
    \and
    NSF NOIRLab, 950 N. Cherry Ave., Tucson, AZ 85719, USA
    \and
    Department of Physics, Southern Methodist University, 3215 Daniel Avenue, Dallas, TX 75275, USA
    \and
    Sorbonne Universit\'{e}, CNRS/IN2P3, Laboratoire de Physique Nucl\'{e}aire et de Hautes Energies (LPNHE), FR-75005 Paris, France
    \and
    Department of Physics \& Astronomy and Pittsburgh Particle Physics, Astrophysics, and Cosmology Center (PITT PACC), University of Pittsburgh, 3941 O'Hara Street, Pittsburgh, PA 15260, USA
    \and
    Department of Physics and Astronomy, University of Waterloo, 200 University Ave W, Waterloo, ON N2L 3G1, Canada
    \and
    Perimeter Institute for Theoretical Physics, 31 Caroline St. North, Waterloo, ON N2L 2Y5, Canada
    \and
    Waterloo Centre for Astrophysics, University of Waterloo, 200 University Ave W, Waterloo, ON N2L 3G1, Canada
    \and
    Departament de F\'isica, EEBE, Universitat Polit\`ecnica de Catalunya, c/Eduard Maristany 10, 08930 Barcelona, Spain
    \and
    Department of Physics and Astronomy, Sejong University, 209 Neungdong-ro, Gwangjin-gu, Seoul 05006, Republic of Korea
    \and
    CIEMAT, Avenida Complutense 40, E-28040 Madrid, Spain
    \and
    Department of Physics \& Astronomy, Ohio University, 139 University Terrace, Athens, OH 45701, USA
    \and
    University of Michigan, 500 S. State Street, Ann Arbor, MI 48109, USA}
    \date{}

  \abstract
    {
    Direct observational constraints on how galaxies acquire their gas remain remarkably limited, hindering our understanding of the baryon cycle.
    We present a search for down-the-barrel \ion{Na}{i} D absorption towards 15.6 million galaxies at $z < 0.6$ in DESI Data Release 2.
    We use Bayesian evidence ratios to assess whether the absorption requires additional components tracing interstellar gas distinct from the systemic component of the galaxy.
    We construct a catalogue of 50 088 galaxies with evidence for down-the-barrel absorption.
    The inferred absorption components are broadly distributed in velocity, with approximately 50\% at $v_{\rm flow} < -50$ \kms, 30\% within 50\kms of the systemic velocity and the remaining 20\% at $v_{\rm flow} > 50$ \kms. 
    We find strong evidence for a large population of low-velocity, infalling absorbers with velocities $\sim$20 \kms in edge-on galaxies, consistent with radial inflows predicted in simulations. 
    The stronger correlation in early-type galaxies between inflow velocity and stellar velocity dispersion, compared to that with stellar mass, suggests that a portion of these inflows may be associated with accreting satellites.
    These results reveal the multiple pathways in which galaxies accrete gas at redshift $z < 0.6$ for the first time in a statistically significant sample.
}
   \keywords{Galaxies: evolution -- Galaxies: active -- Galaxies: ISM -- Galaxies: general -- ISM: jets and outflows 
               }
    \titlerunning{How galaxies accrete cold gas at $z < 0.6$}
    \authorrunning{Weng et al.}
   \maketitle


\section{Introduction}
The formation and evolution of galaxies are fundamentally governed by the availability of gas that fuels star formation. 
The depletion times of molecular gas reservoirs within galaxies are on the order of one Gyr at low redshift and even shorter at higher redshifts \citep{Saintonge2017, Tacconi2018}, implying refuelling of the gas reservoirs to sustain star formation over cosmological timescales.  
To understand how galaxies evolve, it is therefore essential to study the processes that regulate their gas reservoirs, including both the removal of gas through outflows and the replenishment of gas through inflows. 
This continuous exchange, often referred to as the baryon cycle \citep{PerouxHowk2020}, links the intergalactic medium (IGM), the circumgalactic medium \citep[CGM;][]{Tumlinson2017, FaucherOh2023} and the interstellar medium (ISM). 
Gas accretion from the IGM into the CGM and eventually into the ISM \citep{Sancisi2008} is counterbalanced by the expulsion of gas driven by stellar feedback and active galactic nuclei \citep[AGN;][]{Veilleux2005, ThompsonHeckman2024}. 
Observing and disentangling the effects of these gas flows remains one of the central challenges in understanding how galaxies form and evolve. 

Gas accretion has long been recognised as a necessary process for galaxies to sustain star formation \citep{Bouche2010, Dave2012, Lilly2013}.
However, the physical mechanisms by which gas is accreted remain poorly constrained observationally.
At $z \gtrsim 2$ and halo masses $M_{\rm halo} \lesssim 10^{12}$~M$_\odot$, cool gas from the intergalactic medium can free-fall onto galaxies in what is commonly referred to as `cold-mode' accretion \citep[e.g.][]{BirnboimDekel2003, Dekel2006}.
At lower redshifts and in more massive haloes, gas is expected to accrete primarily through the cooling of a hot halo, leading to a smoother inflow onto the disc \citep{Keres2005, Stern2020, Trapp2024}. 
Upon reaching the disc, gas can be channelled inwards via structures such as bars \citep{deV1963, Miller1970}, although observational evidence for this remains scarce \citep{Wong2004, Schmidt2016, DiTeodoroPeek2021}. 
An additional pathway is provided by galactic fountains, in which material ejected by stellar feedback interacts with the hot halo, condenses and subsequently re-accretes onto the galaxy \citep{Shapiro1976, Fraternali2017, Marasco2022}.
Observational evidence for both fountain-driven accretion and cooling from hot haloes has been reported in the Milky Way and nearby galaxies \citep{Marasco2012, Li2023, Sankar2025}. 
Finally, mergers and the accretion of satellites are expected to also play a role \citep{Guo2011}, although this is estimated to account for $\lesssim$ 20\% of inflowing gas \citep{vandeVoort2011, Wang2011, Huillier2012}. 
Distinguishing between these channels remains challenging due to the limited number of direct observations of inflowing gas.

Galactic winds are essential to explain the observed ratio of stellar mass to halo mass. 
At the low mass end, stellar feedback from winds and supernovae explosions suppresses the efficiency with which baryons are converted into stars, while at the high mass end, feedback from active galactic nuclei plays a dominant role \citep{Behroozi2013, Harrison2017}. 
In contrast to accretion, outflows are commonly detected and are observed across all gas phases. 
X ray observations trace the hot ($10^{6\text{-}7}$ K) gas \citep{Strickland2000, Strickland2009}, nebular emission lines trace the warm ($10^{4\text{-}5}$ K) ionised gas \citep{Heckman1990, Sharp2010} and optical and radio absorption or emission lines trace the cooler neutral ($10^{3\text{-}4}$ K) and molecular ($\sim$10 K) gas \citep{vangorkom1989, Cicone2014, Yoon2025FLASH}. 
Although most of the energy in an outflow is carried by the hot gas phase, the bulk of the mass is found in the neutral and molecular components \citep{Fluetsch2021}.  

Many of our observations of galactic winds and gas accretion come from absorption-line probes of neutral and ionised gas against the stellar continuum of the host galaxy, a technique known as down-the-barrel spectroscopy. 
Among the ions targeted in such studies, the \ion{Na}{i} D doublet at vacuum wavelengths $\uplambda\uplambda$5891.583, 5897.558 \AA\ remains one of the most widely used, owing to its accessibility at optical wavelengths at $z \lesssim 0.6$. 
With an ionisation potential of only 5.1 eV, lower than that of \ion{H}{i}, \ion{Na}{i}~D traces the cool, neutral phase of the ISM.

The first down-the-barrel detections of \ion{Na}{i} D absorption were made more than three decades ago in nearby galaxies \citep{Phillips1993}. 
Early studies focused on specific classes of systems, such as luminous infrared galaxies \citep{Heckman2000, Martin2005, Rupke2005aSample, Rupke2005bAnalysis, Cazzoli2016}, before expanding to more typical star-forming and quiescent galaxies \citep{Sato2009, Sun2024}. 
More recently, observations have extended to higher redshift with the advent of the James Webb Space Telescope \citep{Belli2024, Davies2024, Bevacqua2026}. 
Stacking experiments have also been used to measure the average gas-flow properties of galaxies as a function of stellar mass and star-formation rate \citep{ChenSDSS2010, Concas2019, Roberts-Borsani2019}. 
We emphasise, however, that \ion{Na}{i} D is not the only useful tracer: down-the-barrel \ion{Mg}{ii} provides access to higher redshifts \citep{Weiner2009, Erb2012, Martin2012, Rubin2010, Rubin2014}, while UV lines probe warmer ionised gas phases \citep{Steidel2010, Chisholm2015, Heckman2015}.
Despite this wide range of tracers, the field still lacks a truly large sample of individual detections necessary to fully characterise the diversity of gas flows in galaxies.

While outflows are ubiquitously detected across these studies, the corresponding signatures of gas accretion remain remarkably scarce. 
Clear down-the-barrel detections of redshifted inflowing gas are rare and occur with far lower incidence than outflows \citep{Sato2009, Martin2012, Rubin2014}. 
\citet{Sun2024} recently examined the incidence of both outflows and inflows in starburst, post-starburst and quiescent galaxies, finding that only the quiescent population shows predominantly inflowing gas. 
Even in stacking experiments, inflows remain challenging to isolate \citep{Concas2019} and, when tentatively detected, tend to be confined to high-inclination systems \citep{Roberts-Borsani2019}. 
With the advent of integral field spectroscopy (IFS), there has been more recent success in identifying inflows \citep{Roy2021, Rupke2021}, although this has largely been limited to galaxies hosting AGN. 
Radio observations of nearby galaxies have found evidence for \ion{H}{i} inflows \citep[e.g.][]{Wong2004, Schmidt2016, DiTeodoroPeek2021} but the amount of inflowing gas remains debated and sample sizes remain limited. 
More recently, molecular gas inflows at cosmic noon have also been reported \citep{Genzel2023, Jolly2026}.
{
This apparent asymmetry is perhaps expected given the different physical mechanisms governing inflows and outflows. 
Inflow velocities are set by the gravitational potential of the halo and are therefore typically of order the virial velocity, making them difficult to disentangle from internal galaxy kinematics. 
In contrast, outflows are driven by energetic feedback from star formation or AGN activity and can reach velocities that significantly exceed the characteristic dynamical motions of the host galaxy, making them more readily detectable. 
Galaxies must nevertheless be continuously fuelled by gas to sustain star formation, implying that the low observed incidence of inflows reflects observational limitations rather than a lack of accretion itself.}

In this work, we construct a sample of down-the-barrel \ion{Na}{i} D absorbers roughly two orders of magnitude larger than existing samples, using {the currently internal DESI Data Release 2 (DR2) which is intended for a forthcoming public release}.
By identifying absorption features without imposing prior selection on galaxy properties, we aim to build a less biased census of both outflows and inflows than previous, more targeted studies. 
In this first paper, we focus on the methodology used to generate this sample and on the physical interpretation of the resulting absorption features. 
We introduce the DESI survey and describe our parent galaxy sample in \autoref{sec:DESI}.
We then outline the data preparation and Bayesian methods used to identify candidate down-the-barrel absorbers and to derive posterior distributions for the fitted parameters (\autoref{sec:methods}).
\autoref{sec:results} presents measurements of the properties of more than 50 000 detected outflows and inflows, along with the properties of the host galaxies. 
Finally, we discuss the varying origins of these absorbers and outline future applications in \autoref{sec:discuss}. 
Throughout this paper we adopt the cosmological parameters from \citet{Planck2020} {($H_0 = 67.66$ \kms~Mpc$^{-1}$)} and the \citet{Chabrier2003} initial mass function (IMF).

\section{The Dark Energy Spectroscopic Instrument}
\label{sec:DESI}
{
The Dark Energy Spectroscopic Instrument (DESI) is a robotic, fibre-fed, highly multiplexed spectrograph installed on the 4-metre Mayall Telescope at Kitt Peak National Observatory (KPNO) in Arizona \citep{DESI2022.KP1.Instr}. 
Its focal plane contains 5000 robotic fibre positioners \citep{FocalPlane.Silber.2023}, enabling the simultaneous acquisition of spectra from nearly 5000 sources \citep{Corrector.Miller.2023, FiberSystem.Poppett.2024}. 
DESI is conducting an eight-year survey covering $\sim$17\, 000~deg$^2$, and will obtain spectra for approximately 63 million galaxies and quasars, exceeding the initial forecasts of 39 million \citep{DESI2016b.Instr}. 
The scale of the survey necessitates extensive supporting software and survey operations pipelines \citep{Spectro.Pipeline.Guy.2023, SurveyOps.Schlafly.2023}.}

{
Spectroscopic observations began during Survey Validation in 2020 and transitioned to full operations in 2021. 
The first public data release, DESI DR1 \citep{DESI2024.I.DR1}, includes approximately 14.5 million extragalactic spectra and 4 million stellar spectra from the first year of observations. 
Using DR1, cosmological results were obtained from full-shape analyses \citep{DESI2024.VII.KP7B}. 
DESI DR2 expands on this with more than 33 million extragalactic and 12 million stellar spectra from the first three years of the survey \citep{DESI.DR2.BAO.lya, DESI.DR2.BAO.cosmo}, making it one of the most extensive spectroscopic datasets currently available. 
Cosmological analyses are continuing with the upcoming DR3.}

\subsection{The Bright Galaxy Survey}
\label{sec:BGS}
This study uses galaxy spectra from the Bright Galaxy Survey (BGS), one of the five primary DESI target classes. 
The BGS targets galaxies at redshift $z < 0.6$ and is observed under bright-time conditions. 
It consists of two main components: BGS Bright, which includes galaxies with $r \lesssim 19.5$ mag, and BGS Faint, which extends to $19.5 < r < 20.175$ and applies an additional colour–magnitude selection to ensure a high redshift success rate \citep{BGS.TS.Hahn.2023}. 
In this work, we use DESI DR2, which includes approximately 12.2 million galaxies from the BGS sample.
The wavelength coverage of DESI allows robust detection of the \ion{Na}{i} D absorption doublet out to $z \lesssim 0.6$ at moderate spectral resolution ($3800 < R < 5500$). 
This makes the BGS an excellent dataset for identifying and characterising neutral interstellar gas traced by down-the-barrel absorption. 

\subsection{The Luminous Red Galaxy Survey}
\label{sec:LRG}
We also make use of spectra from the Luminous Red Galaxy (LRG) survey, another of the five primary DESI target classes. 
The LRG sample is designed to select intrinsically luminous, massive galaxies out to higher redshift than the BGS, with a primary redshift range of $0.4 < z \lesssim 1.0$ \citep{LRG.TS.Zhou.2023}. 
LRGs are identified through colour–magnitude cuts in optical and infrared imaging, optimised to ensure both high completeness and purity across the redshift range. 
These objects extend our sample at $z < 0.6$ by another 3.4 million. 

\section{Methods}
\label{sec:methods}
We aim to detect and characterise down-the-barrel \ion{Na}{i}~D absorption arising from interstellar gas in a sample of 15.6 million moderate-resolution galaxy spectra. 
This requires several non-trivial steps: establishing a reliable continuum, identifying and modelling absorption components and assessing whether gas flows are detected using the Bayesian evidence. 
In this section, we outline this process and refer readers to Appendices \ref{app:snr5}--\ref{app:PurityCompleteness} for additional details. 

\subsection{Data Preparation}
\subsubsection{Signal-to-noise ratio cut}
We restrict our analysis to galaxies with sufficient signal-to-noise in the \ion{Na}{i}\,D region to allow the detection of down-the-barrel absorption features. 
Specifically, we require a median continuum signal-to-noise ratio (SNR) per pixel $>5$ within a rest-frame $\pm$50\,\AA\ window centred on the \ion{Na}{i}\,D doublet, excluding the doublet itself as well as the nearby \ion{He}{i}\,$\uplambda 5876$ emission line. 
This threshold is comparable to that adopted by \citet{Sato2009} over the redshift range $0.11 < z < 0.54$. 
Including galaxies with $3 < \rm{SNR} < 5$ results in a sample increase of less than 2\%, a finding quantified in \autoref{app:snr5}.
From the initial galaxy sample, we find that roughly $6$ million galaxies meet this threshold. 

\subsubsection{Continuum estimation}
Our goal is to obtain a reliable continuum around the \ion{Na}{i}~D doublet, rather than explicitly modelling and subtracting the stellar absorption component. 
In particular, we do not attempt to separate stellar and interstellar contributions to \ion{Na}{i}~D at the systemic velocity for reasons we discuss in \autoref{app:cont_fit}. 
We begin with \texttt{FastSpecFit}\footnote{https://fastspecfit.readthedocs.io/en/latest/} \citep[][Moustakas et al.in preparation]{MoustakasFastSpecFitsoftware} models, after correcting each spectrum for Galactic reddening using the \citet{Fitzpatrick1999} extinction law. 
The \ion{Na}{i}~D absorption line is interpolated over by masking $\pm 2\sigma_{*, \rm FSF}$ around the doublet, where $\sigma_{*, \rm FSF}$ is the stellar velocity dispersion.
A fixed mask of $\pm400$~\kms is adopted when this is unavailable. 
The \ion{He}{i} emission-line model from \texttt{FastSpecFit} is then added to construct an initial continuum estimate used for normalisation (see the top right of \autoref{fig:Flowchart}).

We further refine the continuum by removing $5\sigma$ outliers outside the masked regions and fitting a Legendre polynomial (orders 1–6), selecting the optimal order via the Bayesian Information Criterion \citep[BIC;][]{Schwarz1978}. 
The final normalised spectrum is obtained by dividing the observed flux by this polynomial. 
As this approach deliberately does not subtract the stellar \ion{Na}{i}~D component, absorption near the systemic velocity may include both stellar and interstellar contributions.

To account for systematic uncertainties in the continuum modelling, we estimate an additional fractional error from the scatter of the flux relative to the combined continuum and emission-line model in the wings of the \ion{Na}{i}~D feature, and add this in quadrature to the spectrum variance. 
We further update modelling uncertainties using a subsample of 1000 galaxies spanning a range of SNR. 
We compare \texttt{FastSpecFit} continua with those derived from the Penalized PiXel-Fitting (\texttt{PPXF}) algorithm \citep{PPXF2004, PPXF2017}, using stellar templates from the X-Shooter spectral library \citep{Verro2022}.
The resulting differences, typically at the few-percent level, are incorporated as an additional wavelength-dependent variance term applied to the full dataset.
We provide further details on the procedures mentioned in \autoref{app:cont_fit}. 

\subsection{Spectral Modelling}
\subsubsection{Model Components}
We model the continuum-normalised \ion{Na}{i}~D doublet using a combination of systemic and gas-flow absorption components. 
The systemic component captures absorption at the galaxy redshift, arising from stellar photospheres, interstellar gas or both. 
As these contributions cannot be reliably separated in individual moderate-SNR spectra, we treat the systemic absorption as a single, phenomenological component.
Additional components offset in velocity trace inflowing or outflowing gas. 
We do not include \ion{Na}{i}~D emission, as it introduces strong degeneracies with absorption in individual spectra.

The systemic absorption is modelled as a double Gaussian with a shared velocity dispersion and centroids near the systemic redshift (\autoref{eq:sys_gauss}). 
Gas flows are described using a partial-covering model \citep[][\autoref{eq:int_flow_phys}]{Rupke2005aSample}, in which the continuum-normalised intensity depends on the covering fraction and optical depth. 
Gaussian optical depth profiles are assumed for each member of the doublet, with the \ion{Na}{i} D oscillator strength ratio fixed to 1:2 (\autoref{eq:opt_depth}). 
This formulation captures the effects of saturation and incomplete covering of the background source. 
The approach follows standard methods in the literature and further details and equations are provided in \autoref{app:models},  \citep{Rupke2005aSample, Rubin2014}.

\subsubsection{Final models}
We construct five models from these components: \texttt{null}, \texttt{sys}, \texttt{flow}, \texttt{sys\_flow}, and \texttt{sys\_flowx2} (see \autoref{tab:models}). 
The \texttt{null}, \texttt{sys}, \texttt{sys\_flow}, and \texttt{sys\_flowx2} models form a nested hierarchy, allowing Bayesian evidence ratios to assess the significance of one or two flow components relative to a baseline model without flows (refer to forthcoming section). 
The \texttt{flow} model tests for purely offset absorption, while the \texttt{sys\_flowx2} model allows for {two} components when required by the data. 
{This limit is appropriate for the typical SNR of our sample and the spectral resolution of DESI, for which additional components would not generally be robustly constrained. 
Post hoc, this is supported by the very small fraction of galaxies requiring two kinematic components ($<2$\%) which we discuss in \autoref{sec:results}, suggesting that systems requiring more than two components are likely to be rarer still.}

\begin{table*}
\centering
\caption{Summary of the five spectral models used to fit \ion{Na}{i} D absorption.}
\label{tab:models}
\begin{tabular}{lllccc}
\hline
\hline
\textbf{Model label} & \textbf{Model description} & \textbf{Equation} &
\textbf{$N_{\mathrm{par}}$} & \textbf{Candidate selection} &
\textbf{Posterior estimation}\\
\hline
\texttt{null} & Null & $K$ & 1 & \checkmark & \ding{55} \\
\texttt{sys} & Systemic & $K \cdot I_{\mathrm{sys}}$ & 5 & \checkmark & \ding{55}\\
\texttt{flow} & Flow & $K \cdot I_{\mathrm{flow}}$ & 5 & \ding{55} & \checkmark \\
\texttt{sys\_flow} & Systemic + Flow &
$K \cdot I_{\mathrm{sys}} \cdot I_{\mathrm{flow}}$ &
9 & \checkmark & \checkmark\\
\texttt{sys\_flowx2} & Systemic + Flow + Flow &
$K \cdot I_{\mathrm{sys}} \cdot I_{\mathrm{flow}} \cdot I_{\mathrm{flow,2}}$ &
13 & \checkmark & \checkmark\\
\hline
\end{tabular}

\tablefoot{
The ``Equation'' column gives the functional form in terms of the component intensities defined in Eqs.~\ref{eq:sys_gauss} and \ref{eq:int_flow_phys}; $K$ is a constant with a value near unity. 
$N_{\mathrm{par}}$ indicates the number of free parameters, described in more detail in Table~\ref{tab:priors}. 
The final two columns indicate whether each model is used for candidate selection or posterior estimation.
}
\end{table*}

\subsubsection{Spectral resolution}
The spectral resolution of DESI varies between $R \sim 2000$ and $R \sim 5500$ \citep{DESI2022.KP1.Instr}. 
At the wavelengths relevant for the \ion{Na}{i}~D doublet, this corresponds to full width at half maximum (FWHM) velocity resolutions of approximately 80~\kms\ at $z = 0$ and 55~\kms\ at $z = 0.6$. 
Before fitting, each absorption model described in the previous section is convolved with the wavelength-dependent resolution matrix provided for every DESI spectrum to ensure that model predictions are compared to the data at the appropriate instrumental resolution.

\subsection{Bayesian Inference and Model Comparison}
We adopt a Bayesian framework to compare models of differing complexity. 
Model comparison is performed using the Bayesian evidence, $\mathcal{Z}$, which naturally penalises additional parameters. 
We quantify the preference for models including gas flows using the difference in log-evidence (also known as the Bayes factor), $\Delta \ln \mathcal{Z}$, and adopt $\Delta \ln \mathcal{Z} > 1$ as the threshold for positive evidence, following the scale of \citet{JeffreysScale}. 
Evidence comparisons are most straightforward when models are nested, as the comparison reduces to whether the improvement in likelihood justifies the increase in prior volume. 
For non-nested models such as \texttt{flow} and \texttt{sys}, the Bayes factor depends more strongly on the choice of priors, making the interpretation of evidence ratios more complex.

\begin{figure*}
    \centering
    \includegraphics[width=0.9\linewidth]{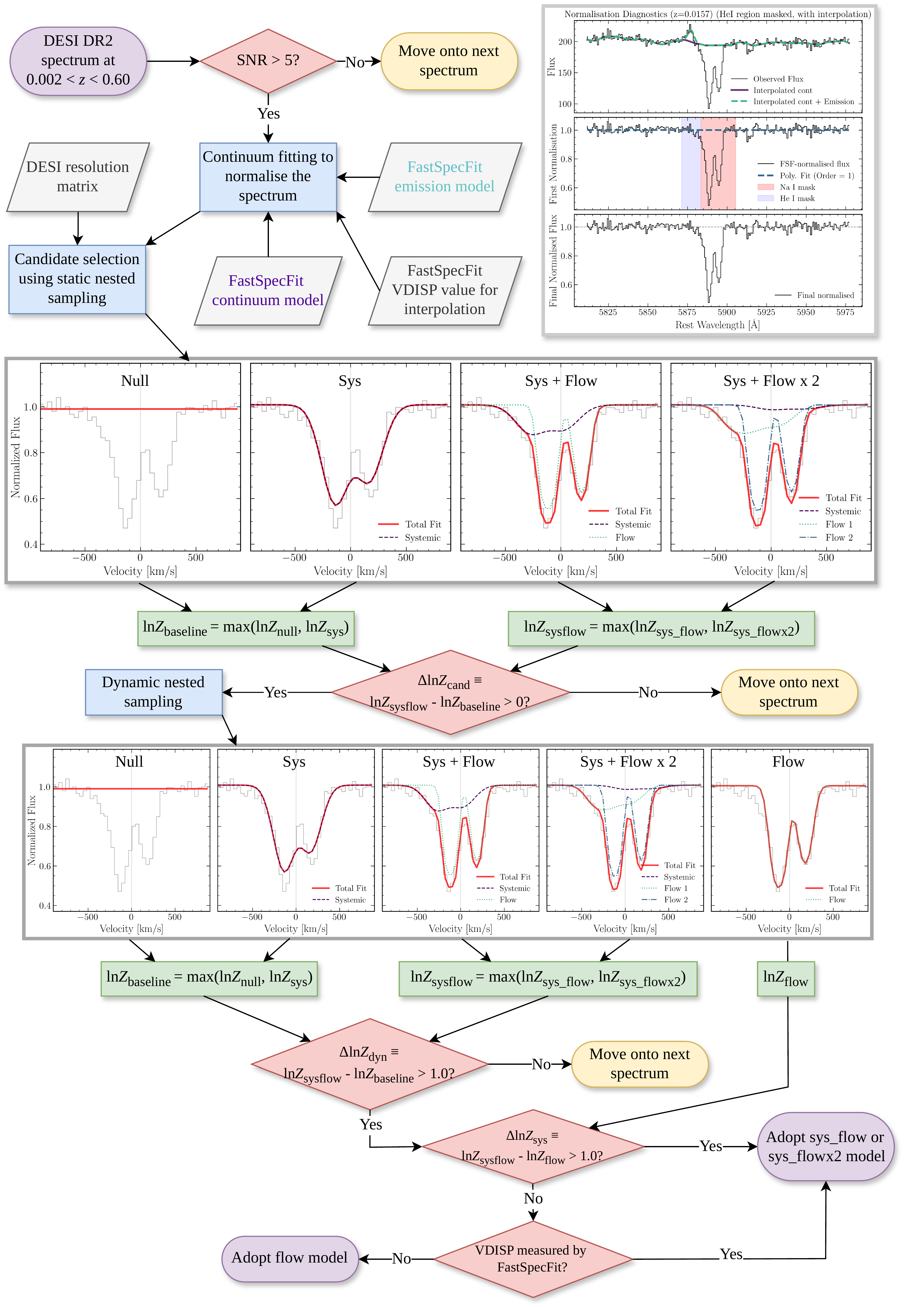}
    \caption{A flow chart illustrating the procedure used to select down-the-barrel candidates. 
    Examples of the continuum subtraction and modelling used for candidate identification and posterior estimation are shown. 
    The subsequent stage, not included in this diagram, is the filtering of the detected absorbers. 
    }
    \label{fig:Flowchart}
\end{figure*}

\subsubsection{Model parameterisation and priors}
We adopt priors that are sufficiently broad to encompass the range of observed down-the-barrel absorption properties, from dwarf galaxies to ULIRGs \citep{SchwartzDwarfOutflow2004, Rupke2005aSample, Martin2005}. 
The systemic velocity is assigned a Student-$t$ prior centred on the \textsc{Redrock} redshift as seen in \autoref{fig:zErrNorm}, incorporating empirical uncertainties from repeat observations \citep{Lan2023, Yu2024} to reduce false detections of low-velocity flows. 

The systemic \ion{Na}{i}~D doublet ratio is restricted to $0.75$–$1.0$, consistent with optically thick absorption \citep{Heckman2000, Rubin2014} and the velocity dispersion is tied to \texttt{FastSpecFit} measurements where available, or otherwise allowed to vary uniformly within 30–450 \kms. 
For gas flows, we adopt a uniform prior on velocity spanning $-1100$ to $1100$ \kms, covering the full range of observed outflows and inflows \citep{Tremonti2007, Rubin2012, Martin2012}. 
We also adopt a uniform prior on the Doppler parameter over 10–500 \kms.
The prior distributions and their values are summarised in  \autoref{tab:priors}, with more detailed justifications found in \autoref{app:zerr}. 

\begin{table*}
\centering
\caption{Prior distributions used for nested sampling.}
\label{tab:priors}
\begin{tabular}{lllp{4cm}p{5cm}}
\hline
\textbf{Component} & \textbf{Parameter} & \textbf{Prior type} &
\textbf{Range} & \textbf{Injected values}\tablefootmark{b} \\
\hline
\multirow{1}{*}{\makecell[lt]{\textit{Null}}}
    & $K$ & uniform & [0.99, 1.01] & 1.0 \\
\hline
\multirow{4}{*}{\makecell[lt]{\textit{Systemic}}}
    & $v_{\rm sys}$ & truncated Student's $t$ &
    see Appendix~\ref{app:zerr} & randomly drawn from real errors \\
    & $EW_{\rm D2}$ & uniform & [0.05, 5.0] & [0.75, 1.0] \\
    & $R_{12}$ & uniform & [0.75, 1.0] & [0.9] \\
    & $\sigma_{\rm sys}$ & uniform &
    $[\;\sigma_{*,\rm FSF} - 3\epsilon_{*,\rm FSF},$\\
    & & & $\;\sigma_{*,\rm FSF} + 3\epsilon_{*,\rm FSF}\;]$
    \tablefootmark{a} & [150, 200] \kms \\
\hline
\multirow{4}{*}{\makecell[lt]{\textit{Flow}}}
    & $v_{\rm flow}$ & uniform & [$-$1100, 1100] \kms &
    [$-$400, $-$250, $-$100, $-$50, 0, 50, 100, 250, 400] \kms \\
    & $C_{f,\rm flow}$ & uniform & [0.0, 1.0] & [0.1, 0.4, 0.7] \\
    & $\tau_{0,D1,\rm flow}$ & uniform & [0.0, 5.0] &
    [0.5, 1.0, 2.0, 4.0] \\
    & $b_{D,\rm flow}$ & uniform & [10, 500] \kms &
    [50, 100, 150, 200, 300] \kms \\
\hline
\end{tabular}
\tablefoot{
Parameters are grouped by the component they describe. Uniform priors are specified by their minimum and maximum values. The Student's-$t$ prior on the systemic velocity depends on the galaxy redshift $z$ and the \texttt{RedRock} redshift uncertainty $z_{\rm err}$ (see Fig.~\ref{fig:zErrNorm}). The final column gives the values used in the completeness tests.
\tablefoottext{a}{Applied only when \texttt{VDISP} is measured by \texttt{FastSpecFit}; otherwise, a uniform prior over [30, 450]~\kms is used.}
\tablefoottext{b}{A model without a systemic component is also injected in the completeness tests.}
}
\end{table*}

\subsubsection{Stage I: Candidate pre-selection}
In Stage I, our aim is to rapidly identify a highly complete set of potential down-the-barrel absorbers using a deliberately permissive Bayes factor threshold. 
We use nested sampling \citep{Skilling2004, Skilling2006}, as implemented in \texttt{dynesty} \citep{SpeagleDYNESTY, DYNESTY2025}, to compute Bayesian evidences for the four models in \autoref{tab:models}.

For each galaxy, we evaluate these nested models, divided into (i) baseline models without gas flows (\texttt{null}, \texttt{sys}) and (ii) flow models with one or two components (\texttt{sys\_flow}, \texttt{sys\_flowx2}). 
Candidate selection is based on the relative evidence, $\Delta \ln \mathcal{Z_{\rm cand}} = \ln \mathcal{Z}_{\rm sysflow} - \ln \mathcal{Z}_{\rm baseline}$, where each term corresponds to the preferred model within its class.

For efficiency, we use a stopping criterion of $\texttt{dlogz} = 0.01$ and $n_{\rm live} = 20 \times n_{\rm dim}$ live points. 
Although this choice is modest \citep{FerozHobson2019}, we validate it in \autoref{app:nest_cand} by comparing evidence values from this run with those from a longer dynamic nested sampling run on a randomly selected subsample of $\sim$20{,}000 galaxies.
To maximise completeness, we apply a permissive threshold of $\Delta \ln \mathcal{Z_{\rm cand}} > 0$, ensuring that spectra with even marginal support for a flow component are retained. 
This produces an intentionally impure but highly complete sample for Stage II.

\subsubsection{Stage II: Final model selection and posterior estimation}
Following Stage I, we retain $\sim$110,000 galaxies. 
In Stage II, we re-run the inference using dynamic nested sampling in \texttt{dynesty}, which improves evidence and posterior estimations. 
We initialise with 500 live points, allow additional batches of 100 and adopt a stopping criterion of \texttt{dlogz} $= 0.01$ with posterior weighting of 0.9. 
Parameter estimates are taken as the median of the posterior, with uncertainties given by the 16--84th percentiles.

We define our primary sample by requiring $\Delta \ln \mathcal{Z_{\rm dyn}} > 1.0$, yielding $\sim$65,000 galaxies with statistically significant down-the-barrel absorption. 
For these systems, we select between \texttt{sys\_flow} and \texttt{sys\_flowx2}, retaining the second component only if $\Delta \ln \mathcal{Z_{\rm comp}} > 1.0$.

For galaxies lacking reliable stellar velocity dispersions, we additionally compare \texttt{sys\_flow} and \texttt{flow}. 
The systemic component is retained only if $\Delta \ln \mathcal{Z_{\rm sys}} > 1.0$, otherwise the \texttt{flow} model is adopted. 
This allows for systems with purely offset absorption. 
The full procedure is summarised in \autoref{fig:Flowchart}, with evidence thresholds summarised in \autoref{tab:dlogz_meaning}.

\begin{table*}
\centering
\caption{Definitions of the evidence comparisons used in this work.}
\label{tab:dlogz_meaning}
\begin{tabular}{llll}
\hline\hline
\textbf{Purpose} & \textbf{Lower-$n_{\rm dim}$ model} &
\textbf{Higher-$n_{\rm dim}$ model} & \textbf{Threshold} \\
\hline
Candidate selection using static nested sampling &
BEST(\texttt{null}, \texttt{sys}) &
BEST(\texttt{sys\_flow}, \texttt{sys\_flowx2}) &
$\Delta\ln\mathcal{Z}_{\rm cand}>0.0$ \\

Final selection using dynamic nested sampling &
BEST(\texttt{null}, \texttt{sys}) &
BEST(\texttt{sys\_flow}, \texttt{sys\_flowx2}) &
$\Delta\ln\mathcal{Z}_{\rm dyn}>1.0$ \\

Check if the flow-only model is preferred &
\texttt{flow} & \texttt{sys\_flow} &
$\Delta\ln\mathcal{Z}_{\rm sys}>1.0$ \\

Check if the second flow component is needed &
\texttt{sys\_flow} & \texttt{sys\_flowx2} &
$\Delta\ln\mathcal{Z}_{\rm comp}>1.0$ \\
\hline
\end{tabular}
\tablefoot{
In the following sections, we use
$\Delta\ln\mathcal{Z}\equiv\Delta\ln\mathcal{Z}_{\rm dyn}$
for brevity.
}
\end{table*}

\subsection{Candidate filtering}
The final step of the pipeline is to filter the remaining $\sim$65,000 candidates to remove clear non-astrophysical false positives. 
Based on visual inspection of several thousand spectra and their residuals, we identify recurring spurious detections and apply a set of empirically calibrated cuts to produce a cleaner sample.

We exclude components with very narrow widths ($b_D < 20$ \kms), which are typically driven by spectral artefacts, and remove extreme velocity outliers ($v_{\rm flow} < -900$ \kms\ or $v_{\rm flow} > 700$ \kms) that show low purity. 
Shallow features are suppressed by requiring a maximum absorption depth exceeding $1.5\sigma$ and foreground Galactic \ion{Na}{i}~D contamination is removed by excluding systems within $|v_{\rm flow} - v_{\rm MW}| < 150$ \kms. 
We also inspect objects classified as \texttt{PSF} in DR9 Legacy Survey imaging, finding a $\sim$17\% contamination rate from stars or quasars. 
Finally, catastrophic fits are removed using a dynamic $\ln \mathcal{Z}$ threshold, defined from the lowest 1\% of values in SNR bins, which rejects poorly constrained fits and spectra with strong \ion{Na}{i}~D emission.

These criteria are applied on a component-by-component basis, such that invalid components are removed and \texttt{sys\_flowx2} systems may be reduced to \texttt{sys\_flow}. 
We do not modify the priors, which are intentionally broad and physically motivated, but instead apply conservative post-selection cuts to account for modelling limitations.
This filtering prioritises purity over completeness and may bias against certain outflow morphologies, particularly broad, shallow features that are difficult to distinguish from continuum uncertainties. 
However, given that genuine down-the-barrel systems comprise only $\sim$0.8\% of the parent sample, even a small contamination fraction would dominate the final catalogue. 
We discuss future improvements in \autoref{sec:discuss}.

\subsection{Completeness and purity}
We assess our detection pipeline through a combination of visual inspection and injection–recovery tests. 
Visual vetting indicates a low rate of direct false positives ($<3$\%), while our Bayesian model selection framework further controls statistical false detections through a conservative evidence threshold. 
Injection tests show that completeness depends strongly on SNR and absorption properties, with higher recovery rates for high-velocity and narrow features compared to broad or near-systemic components. 
We refer readers to \autoref{app:PurityCompleteness} for a more quantitative discussion. 

\section{Results}
\label{sec:results}
For our default evidence threshold of $\Delta \ln \mathcal{Z}_{\rm dyn} > 1.0$ (hereafter $\Delta \ln \mathcal{Z}$), we identify {50 088} galaxies with \ion{Na}{i}~D absorption consistent with a down-the-barrel geometry.
A small fraction of the sample contains multiple gas-flow components, with 747 galaxies requiring two components, corresponding to 50 835 absorption components in total. 
For the $\Delta \ln \mathcal{Z} > 1.0$ sample, 18.0\% of galaxies are best described by a flow-only model and therefore do not require a systemic component.

{As the stringency of the selection is increased, the sample size decreases to 27 420 galaxies for $\Delta \ln \mathcal{Z} > 3.0$ and 15 696 galaxies for $\Delta \ln \mathcal{Z} > 5.0$, corresponding to 28 112 and 16 287 absorption components, respectively. 
The fraction of flow-only galaxies also decreases with increasing evidence threshold, from 18.0\% at $\Delta \ln \mathcal{Z} > 1.0$ to 13.0\% and 4.2\% for the $\Delta \ln \mathcal{Z} > 3.0$ and $5.0$ samples, respectively. 
The main results are generally unchanged when adopting these higher-evidence samples. 
However, increasingly stringent evidence thresholds are biased towards narrower absorption features, which can be more easily distinguished from the systemic component. }
All figures and discussion will use the $\Delta \ln \mathcal{Z} > 1.0$ sample.

Compared to previous individual surveys of down-the-barrel absorption, regardless of the specific ion species, which have typically been limited to several hundred detections, this dataset represents a two-order-of-magnitude increase in sample size. 
This provides an unprecedented sample for understanding how gas flows affect the evolution of galaxies. 

\subsection{Identifying outflows and inflows}
Throughout this work, velocities refer to the line-centre velocity, defined as the velocity corresponding to the minimum flux of the absorption profile. 
For the systemic reference frame, one may use either the \ion{Na}{i}~D systemic component recovered by the nested-sampling fits or the \texttt{RedRock} pipeline redshift.  
We adopt the fitted systemic component because it is a self-consistent reference for measuring velocity offsets.

\begin{figure}
    \centering
    \includegraphics[width=\linewidth]{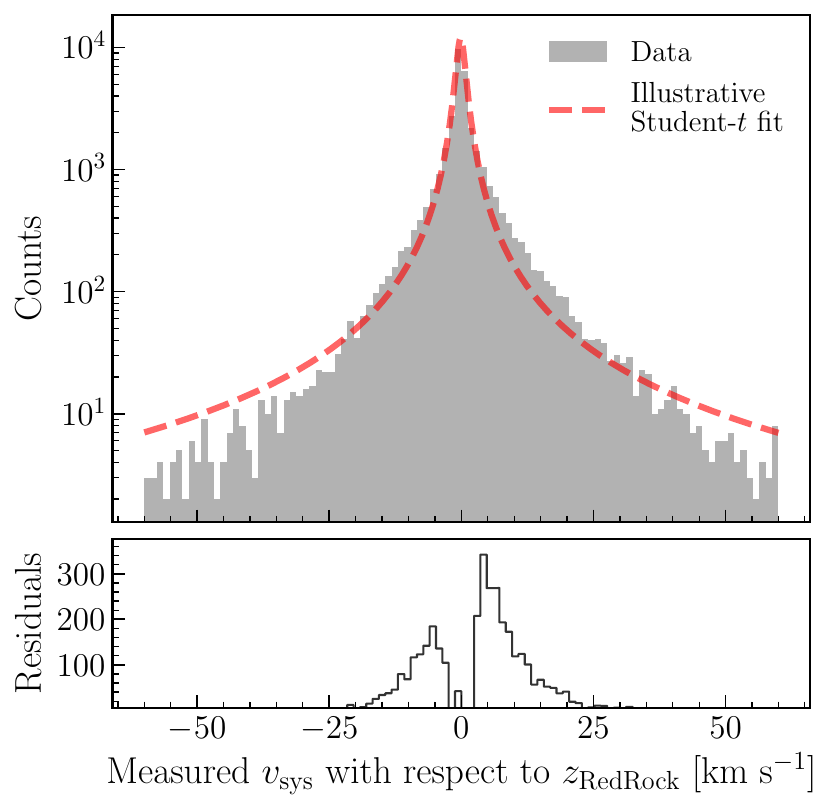}
    \caption{
        Distribution of velocity differences between the systemic component inferred from nested sampling and the \texttt{RedRock} redshift.
        A representative Student-$t$ model (red dashed line) is overplotted for illustration. 
        The difference between the fit and the observed differences is shown in the bottom panel.  
    }
    \label{fig:z_err_dist}
\end{figure}

In \autoref{fig:z_err_dist}, we show the distribution of velocity differences between the fitted systemic component and the corresponding \texttt{RedRock} redshift.
Large discrepancies are rare, with the counts falling below $\sim$10 per 1 \kms bin for velocity offsets $\gtrsim 30$~\kms.
We overlay a representative Student-$t$ fit, noting that it is not expected to be exact because the adopted priors depend on redshift, galaxy type (BGS or LRG) and the \texttt{ZERR} output from \texttt{RedRock}.
Nevertheless, the residuals reveal a significant excess at small positive velocities ($\sim$5--20~\kms), indicating that the fitted systemic component is, on average, slightly redshifted relative to the pipeline value.
This excess is asymmetric, being larger than at comparable negative velocities: there are $723$ galaxies with residual velocities in the $+10$ to $+20$~\kms window compared to $297$ at $-20$ to $-10$~\kms.
To quantify this asymmetry, we apply a binomial test under the null hypothesis that, for galaxies falling in either window, each has equal probability of landing on the positive or negative side.
The resulting p-value is $p \approx 0$, allowing us to reject the null hypothesis of symmetry at high significance.
We interpret this feature as evidence that slow inflowing gas is partially absorbed into the systemic component during the fitting process, an effect explored in more detail in the following sections.

A velocity threshold of $50$~\kms\ is commonly adopted in the literature to separate gas near the systemic redshift from inflowing or outflowing components. 
Strictly speaking, inflow implies that the gas is gravitationally bound to the host galaxy and outflow implies the gas was (or perhaps still is) gravitationally bound; this is not necessarily the case in the sample of absorbers presented here. 
We find that objects with inflow velocities $>200$~\kms\ are often late-stage mergers, where the apparent ``inflow'' may arise from the systemic component of the infalling galaxy rather than from gas gravitationally bound to the primary system. 
In addition, galaxies can function as background sources that probe the discs of foreground galaxies at small transverse separations. 
We will discuss these various scenarios in more detail in \autoref{sec:discuss}. 
We still divide the sample into low-velocity components, defined as $|v| \leq 50$~\kms, and high-velocity components, defined as $|v| > 50$~\kms.
For high-velocity absorbers we further distinguish blueshifted ($v < -50$~\kms) and redshifted ($v > 50$~\kms) gas.
Throughout this work, we use the terms ``outflow'' and ``inflow'' loosely to refer to blueshifted and redshifted gas respectively, without implying a specific physical phenomenon unless directly stated.

Applying these criteria to our default sample ($\Delta \ln \mathcal{Z} > 1.0$), we find that 24 767 components (48.7\%) are classified as outflows ($v_{\rm flow} < -50$ \kms), while 11 333 (22.4\%) are identified as inflows ($v_{\rm flow} > 50$ \kms). 
The remaining 14 735 components (34.6\%) reside within $\pm 50$ \kms of the systemic velocity and are classified as low-velocity. 
As the evidence threshold is increased, the relative proportions of these kinematic classes shift, as summarised in Table \ref{tab:velocity_stats}. 
Notably, the proportion of low-velocity components increases from 29.0\% at $\Delta \ln \mathcal{Z} > 1.0$ to 43.8\% at $\Delta \ln \mathcal{Z} > 5.0$. 
These highest-confidence detections in our sample are associated with narrow features near systemic velocity rather than the typically broader absorption tracing the bulk motions of the gas.

\begin{table*}[h]
\centering
\caption{Distribution of kinematic components across different Bayesian evidence thresholds.}
\label{tab:velocity_stats}
\begin{tabular}{lcccccc}
\hline
\textbf{Threshold} & \textbf{\begin{tabular}{@{}c@{}}Total\\Galaxies\end{tabular}} & \textbf{\begin{tabular}{@{}c@{}}Total\\Components\end{tabular}} & \textbf{Outflows} & \textbf{Low-velocity} & \textbf{Inflows} & \textbf{Flow-only} \\
 & & & ($v_{\rm flow} < -50$ \kms) & ($|v_{\rm flow}| \le 50$ \kms) & ($v_{\rm flow} > 50$ \kms) & \\
\hline
$\Delta \ln \mathcal{Z} > 1.0$ & 50 088 & 50 835 & 24 767 (48.7\%) & 14 735 (29.0\%) & 11 333 (22.3\%) & 9032 (18.0\%) \\
$\Delta \ln \mathcal{Z} > 3.0$ & 27 420 & 28 112 & 11 835 (42.1\%) & 9867 (35.1\%) & 6410 (22.8\%) & 3577 (13.0\%) \\
$\Delta \ln \mathcal{Z} > 5.0$ & 15 696 & 16 287 & 5054 (31.0\%)  & 7130 (43.8\%)  & 4103 (25.2\%) & 663 (4.2\%) \\
\hline
\end{tabular}
\end{table*}

\begin{figure*}
\centering
\includegraphics[width=\linewidth]{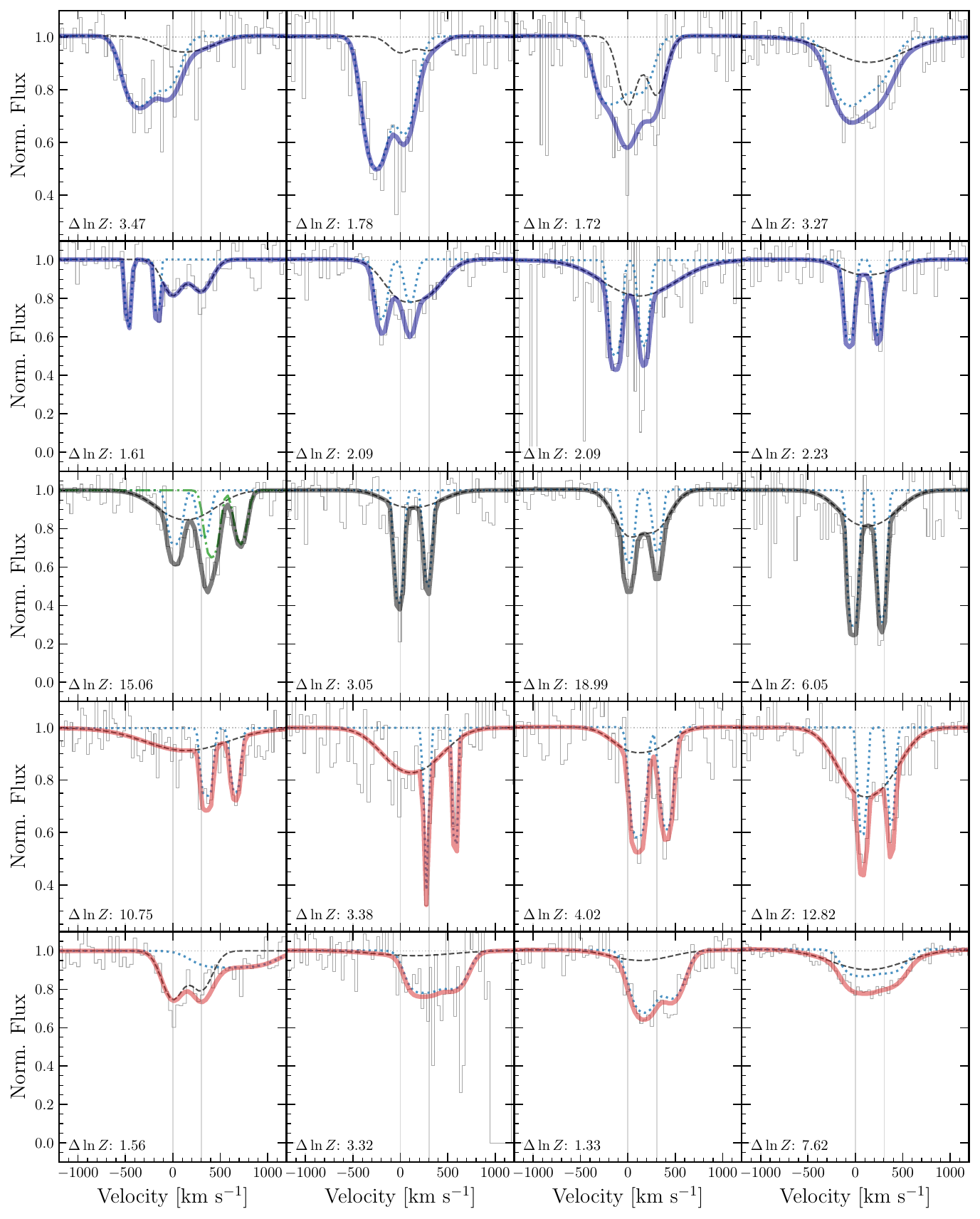}
\caption{
    Representative examples of down-the-barrel \ion{Na}{i} D absorption with $\Delta \ln \mathcal{Z} > 1.0$.  
    Rows correspond to different kinematic classes: outflows (top two rows), low-velocity components (middle row in black) and inflows (bottom two rows).  
    Within each class, spectra are ordered from left to right by decreasing absolute velocity offset from the systemic redshift.  
    The outer rows of the outflow and inflow subsets contain broad absorbers ($b_D > 100$ \kms), whereas the inner rows show narrower components ($b_D < 50$ \kms).  
    Fitted models include both the \texttt{flow} and \texttt{sys\_flow} solutions: the systemic component in the latter is shown as a dashed purple line and the flow component as a dotted cyan line. 
    Vertical grey lines mark the expected locations of the \ion{Na}{i} D doublet at the galaxy’s systemic redshift.
}
\label{fig:examples}
\end{figure*}

To illustrate the diversity of kinematic signatures recovered by our modelling, we present in \autoref{fig:examples} a set of 20 representative down-the-barrel absorption candidates.
The examples are arranged by kinematic class: the top two rows show outflowing gas (blue), the middle row displays components consistent with the systemic velocity (grey) and the bottom two rows show inflowing gas (red).
Within each class, the left to right ordering reflects decreasing absolute velocity offset from systemic.
To highlight the range of Doppler widths present in the sample, the outer top and bottom rows contain systems with broad components ($b_D > 100$ \kms), while the immediately adjacent rows show narrower features ($b_D < 50$ \kms).
For each example, we overplot both the systemic and gas flow model components (if both are required, otherwise only gas flow): the systemic absorption is shown as a dashed purple curve and the flow component as a dotted cyan curve.
These examples collectively demonstrate the range of line shapes and velocity offsets that our method is able to isolate with high statistical confidence.

\subsection{Distribution of galaxy and measured outflow properties}
To characterise the host galaxies of our down-the-barrel absorption sample, we adopt stellar masses from a value-added catalogue derived using the Code Investigating GALaxy Emission \citep[CIGALE;][]{BurgarellaCIGALE2005, NollCIGALE2009, Boquien2019}.
This catalogue, prepared for DESI DR2, is not yet publicly released; for methodological details, we refer readers to the earlier public implementation on the Early Data Release \citep{ZouCIGALE2024}.
CIGALE provides key physical properties, including stellar mass, which form the basis for our analysis of how inflows and outflows correlate with galaxy properties.

\begin{figure*}
\centering
\includegraphics[width=\linewidth]{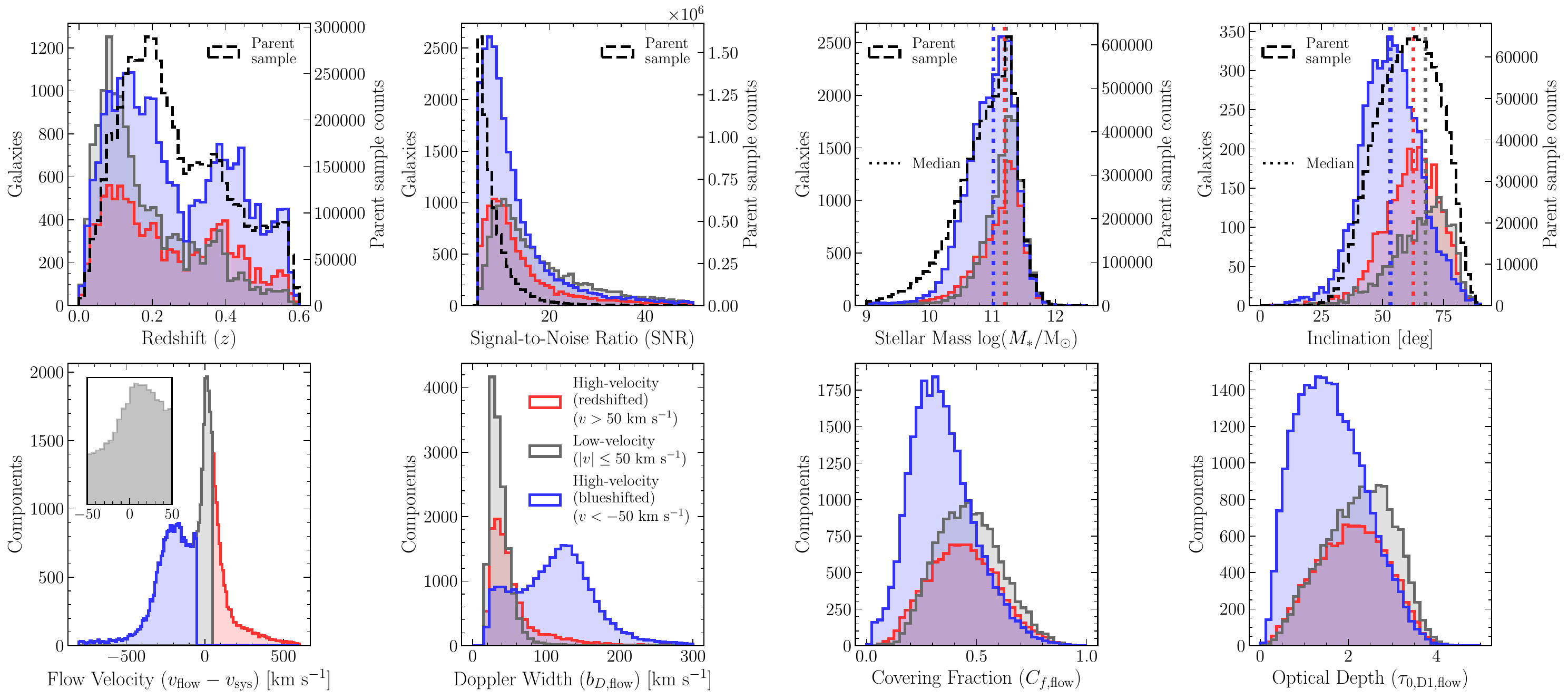}
\caption{Distribution of galaxy properties and their measured gas flow parameters. 
        In the top row, we show from left to right, the galaxy redshift, signal-to-noise ratio of the spectra, galaxy stellar mass and inclination. 
        On the bottom, we show the distribution of measured gas flow velocities with respect to systemic, Doppler widths, covering fractions and optical depths. 
        The dashed histograms show the distribution of properties for galaxies with ${\rm SNR} > 5$ comprising the parent sample where we searched for down-the-barrel absorption. 
        The dotted vertical lines correspond to median stellar masses and inclinations in the two upper-right panels.  
        }
\label{fig:param_dist}
\end{figure*}

We estimate galaxy inclinations using shape parameters from the Legacy Survey DR9 Tractor catalogue \citep{Dey2019}.  
Measured ellipticities are converted into apparent axis ratios, from which the inclination $i$ is computed assuming an intrinsic axis ratio of $q_0 = 0.13$ \citep{Giovanelli1994}. 
Inclinations are calculated for the subset of 6179 disc galaxies with \texttt{EXP} (exponential) or $n < 2$ \texttt{SER} (S\'{e}rsic) profiles, for which this approximation is most appropriate \citep{Blanton2009}.
With these galaxy properties in hand, we present the distributions of the measured galaxy and gas-flow parameters in \autoref{fig:param_dist}.  
The top row summarises properties of the host galaxies, while the bottom row shows the inferred kinematic and physical parameters of the gas-flow components.  

\subsubsection{Redshift and SNR distributions}
The redshift distribution of galaxies with detected down-the-barrel absorption does not follow that of the full galaxy sample with continuum SNR $>5$ around \ion{Na}{i}~D, as shown by the filled and dashed histograms in the upper-left panel of \autoref{fig:param_dist}. 
We find an excess of detections at low redshift, driven by the fact that our completeness increases with SNR and nearby galaxies appear brighter. 
The adjacent panel shows the corresponding SNR distribution: although four times as many galaxies have SNR $\sim$5 compared with SNR $\sim$10, the number of detections peaks around SNR $= 10$. 
Together, these trends simply reflect that our detection sensitivity improves with increasing brightness of the background stellar continuum.

\subsubsection{Absorption traces massive galaxies}
Using the stellar masses estimated by CIGALE, we find that down-the-barrel absorption is preferentially detected in massive galaxies. 
Only 2.3\% of the sample has a stellar mass $\log (M_*/\mathrm{M_\odot}) < 10.0$ and the median stellar mass is $\log (M_*/\mathrm{M_\odot}) = 11.0$ (11.2) for galaxies hosting outflowing (inflowing and systemic) components. 
This trend is consistent with \ion{Na}{i}~D stacking experiments \citep{ChenSDSS2010, Concas2019, Roberts-Borsani2019}, which show that significant excess interstellar absorption emerges primarily in stacks of galaxies with $\log (M_*/\mathrm{M_\odot}) \gtrsim 10.5$. 
Such behaviour is expected: massive galaxies have brighter stellar continua, increasing our sensitivity to absorption signatures. 
Furthermore, the low ionisation potential of \ion{Na}{i}~D makes it particularly abundant in dusty systems, which themselves tend to be more massive.

\subsubsection{Inclination}
Galaxy inclination has a well-established influence on the detectability and measured properties of down-the-barrel absorption. 
Outflows are often observed to escape preferentially along the minor axis and are therefore more readily detected in face-on (low-$i$) systems. 
Conversely, inflowing gas has been reported to occur more frequently in edge-on galaxies \citep{Rubin2012, Rubin2014, Roberts-Borsani2019}, although lower inclinations might be expected if some inflows arise from galactic fountains rather than accretion parallel to the disc. 

{
The inclination distributions of our subsamples in the top-right panel of \autoref{fig:param_dist} are broadly consistent with these expectations, with outflows skewed towards lower inclinations and inflowing or low-velocity components more prevalent at higher inclinations. 
We present a more quantitative analysis of the inclination dependence in \autoref{sec:discuss}, where we examine these distributions in detail.}

\subsubsection{Low-velocity absorbers near systemic velocity are not tracing bulk ISM kinematics}
The components we classify as ``low-velocity'' (those within $\pm 50$ \kms of $v_{\rm sys}$) are distinct from the systemic component. 
As already mentioned, the systemic model in our fits captures the combination of stellar \ion{Na}{i} D absorption and any associated interstellar absorption with widths comparable to the stellar velocity dispersion. 
In contrast, the flow components identified near $v_{\rm sys}$ have much narrower Doppler widths ($b_D \ll \sigma_*$) and therefore cannot arise from the same kinematic structure. 
These narrow, near-systemic features may represent low-velocity instances of other physical processes, rather than the rotating gas in the ISM. 
We discuss this possibility in more detail in \autoref{sec:discuss}. 

\subsubsection{Bimodality in Doppler widths}
The distribution of Doppler widths ($b_D$) for the gas-flow components exhibits clear structure rather than a single, smooth population.
In particular, we observe evidence for two characteristic regimes: a population of relatively narrow components ($b_D \lesssim 70$ \kms) and a broader population extending to several hundred \kms in line width.
This bimodality suggests that we are probing gas with distinct kinematic and physical properties.

The narrow components are indicative of less turbulent gas and are dominated by absorbers near systemic velocity or infalling, while the broader components trace more dynamically disturbed material associated with outflows. 
While there is a substantial population of narrow-line outflows, the fraction of broad inflowing or near-systemic absorbers is considerably smaller. 
Part of this imbalance is observational: by treating the systemic stellar and interstellar absorption jointly, our sensitivity to gas components with line widths comparable to the stellar velocity dispersion is reduced, making broad components near $v_{\rm sys}$ more difficult to recover. 

{
However, this difference is also likely to reflect the underlying physics of the gas flows. 
Inflowing gas is expected to have relatively low internal velocity dispersions set by gravitational infall and modest turbulence, whereas outflows are driven by energetic feedback from star formation or AGN activity, which can generate highly turbulent gas and hence broader line profiles. 
The relative scarcity of broad inflowing components therefore suggests that highly turbulent inflowing gas is intrinsically uncommon, rather than simply missed due to observational limitations.}

\subsubsection{Covering fraction and optical depth}
The inferred covering fractions and optical depths of the gas-flow components span a wide dynamic range (\autoref{fig:param_dist}).
The covering fraction distribution peaks at $C_f \sim 0.4$, with relatively few systems consistent with complete coverage.
This behaviour indicates that the absorbing gas typically does not uniformly cover the stellar continuum, instead favouring a patchy geometry along the line of sight.
The optical depth distribution similarly extends from optically thin to saturated regimes.
The large fraction of components with $\tau > 1$ indicates that the absorbing gas is generally optically thick, consistent with past observations \citep{Rupke2005aSample, Martin2005}.

We caution that the inferred covering fraction and optical depth are not independent parameters.
These quantities are partially degenerate: deeper absorption can be produced either by increasing the optical depth or by increasing the covering fraction {(as described in \autoref{app:tau_degen})}.

\subsection{Morphology}
In addition to inclination, we investigate the morphological properties of galaxies hosting down-the-barrel absorption. 
Early-type galaxies (ETGs) are selected from the Legacy Survey DR9 using the \texttt{DEV} (de Vaucouleurs) and \texttt{SER} classifications, with the latter required to have S\'{e}rsic index $n_{\rm SER} > 3$. 
As before, we define disc galaxies as those with $n_{\rm SER} < 2$.
{This leaves a subset of galaxies with $2 < n_{\rm SER} < 3$ that remain unclassified, as it is more ambiguous whether they are discs or ETGs. }
Disc galaxies are selected following the criteria described previously and we include a third category comprising objects classified as \texttt{REX} (round exponential) or \texttt{PSF}. 
The \texttt{REX} class corresponds to slightly extended, typically low-SNR sources, while galaxies labelled \texttt{PSF} are similarly compact. 
As noted earlier, all \texttt{PSF}-classified objects have been visually inspected {spectroscopically} to remove contamination from stars or quasars and those retained in the final sample represent genuine galaxies exhibiting down-the-barrel absorption.

\begin{figure*}
    \centering
    \includegraphics[width=\linewidth]{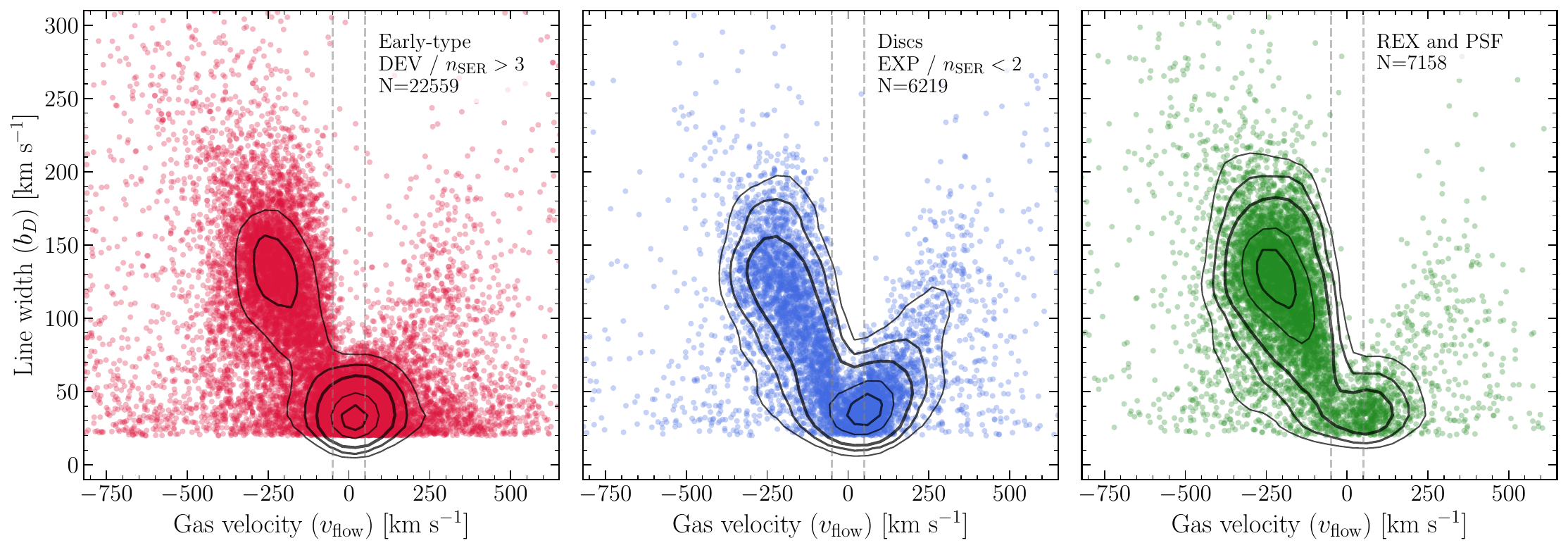}
    \caption{
    Down-the-barrel absorption properties for galaxies of different morphologies. 
    We divide the sample into early-type, disc and compact galaxies (from left to right) using the Legacy Survey DR9 Tractor photometry catalogue. 
    Contours show the 5th, 10th, 20th, 50th and 80th percentiles, and the number of galaxies belonging to each class is indicated in the top right. 
    {In total, only 35 936 galaxies are represented here, due to both missing Legacy Survey classification and the difficulty of classifying galaxies with $2 < n_{\rm SER} < 3$ as either early-type galaxies or discs.}}
    \label{fig:morph}
\end{figure*}

In \autoref{fig:morph}, we show the typical line widths and velocities of absorbers associated with these three morphological classes. 
The contours highlight the regions occupied by most of the population. 
In the left panel, early-type galaxies dominate the absorber-host population, accounting for 63\% of classified systems. 
{This early-type fraction is broadly consistent with the target selection of the final sample: 28 571/50 088 galaxies (57.0\%) are BGS galaxies and 21 517/50 088 (43.0\%) are LRG galaxies, noting that some BGS galaxies are also likely to be early-type systems.}
Most absorption in ETGs lies close to systemic velocity and exhibits narrow line widths. 
The contours are not symmetric about zero: their centres favour small positive velocities, consistent with the trends seen in \autoref{fig:param_dist}. 
A notable tail toward negative velocities with higher $b_D$ also indicates the presence of broader outflows. 
Although we do not measure star-formation rates here, the prevalence of such features in ETGs, typically less star-forming than discs, suggests a significant population of outflows in more quiescent galaxies, consistent with previous findings \citep{Sato2009, Sun2024}. 

Turning to disc galaxies (central panel), we again find a strong concentration of absorbers with low $|v_{\rm flow}|$ and $b_D < 50$ \kms. 
The distribution is more asymmetric than for ETGs, with its centre shifted to approximately $+50$ \kms, indicative of a substantial infalling population. 
A larger fraction of disc galaxies also host outflows and the pronounced tail towards positive velocities and larger $b_D$ suggests that broader inflows are more common in this morphological class.

Finally, galaxies classified as \texttt{REX} and \texttt{PSF} (right panel) appear to be dominated by outflows. 
This is unsurprising, as these compact classifications often include pea galaxies with extremely high specific star-formation rates \citep{Cardamone2009}.

\section{Discussion}
\label{sec:discuss}
Using the measured gas-flow properties, we then examine how they correlate with key galaxy parameters such as redshift, inclination and stellar mass.  
We also present evidence for a population of weakly redshifted absorbers that fall below our nominal inflow threshold of $+50$~\kms\ but nonetheless appear to trace the same physical processes as higher-velocity inflows.  
These observational trends provide the foundation for the subsequent discussion on the physical origins of the absorbers in this sample.

\subsection{Variation of gas flow properties with redshift, inclination and stellar mass}
Interpreting how gas-flow properties vary with galaxy properties requires care, because several observational and physical effects change systematically with redshift.  
Most importantly, the physical area subtended by the 1.5 arcsec DESI fibre varies by more than two orders of magnitude across our sample: from $\lesssim 100$~pc at very low redshift to nearly 10~kpc by $z = 0.6$.  
This strongly affects which regions of a galaxy are sampled.  
In addition, our SNR threshold and methods bias us towards more massive galaxies at higher redshift.  
Cosmic evolution may also play a role: $z = 0.6$ corresponds to a lookback time of $\sim$6~Gyr, over which rates of star formation, AGN activity and gas cycling are all expected to evolve.

\begin{figure*}
    \centering
    \includegraphics[width=\linewidth]{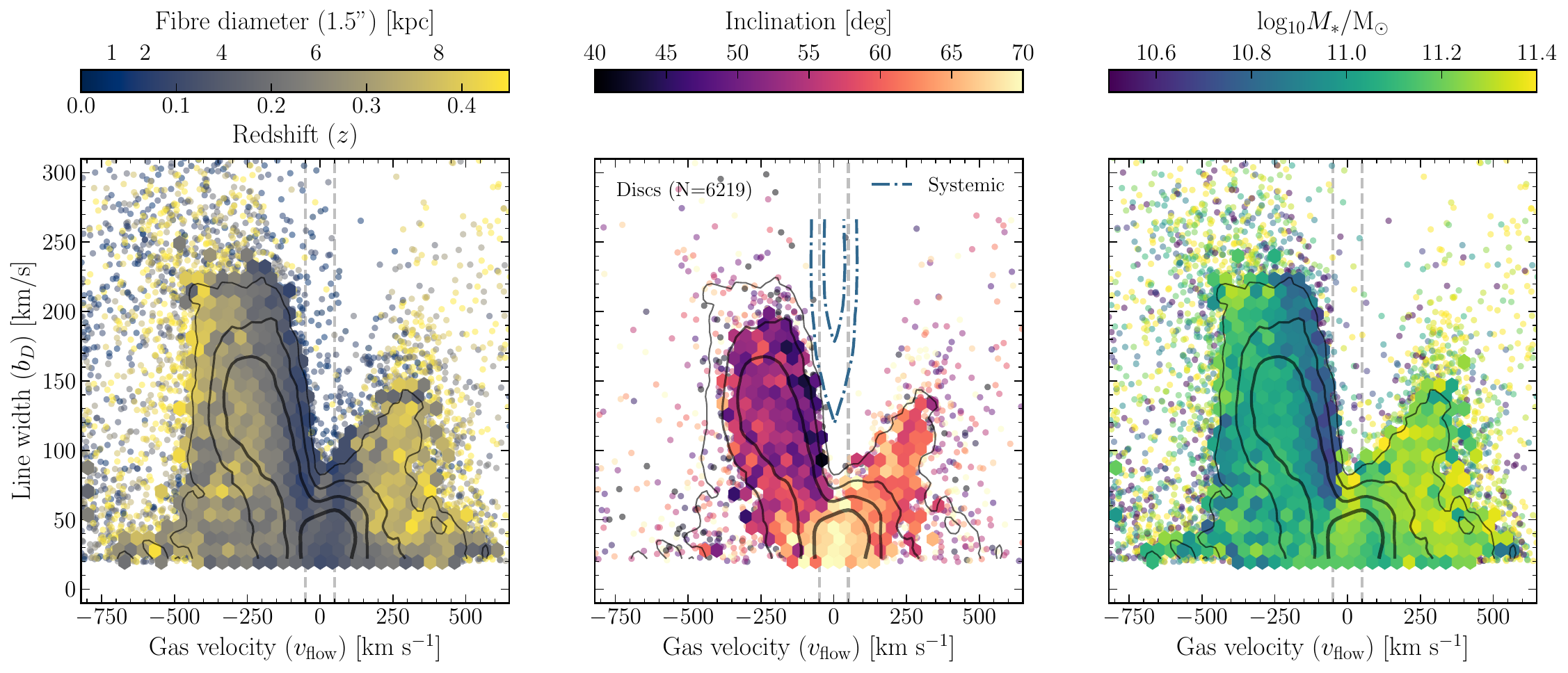}
    \caption{    
    The kinematic phase space of \ion{Na}{i} D absorption, showing Doppler width ($b_D$) versus flow velocity ($v_{\rm flow}$).
    Hexagonal bins are coloured by the median galactic property within each bin, with individual points overplotted in low-density regions. 
    Black contours indicate the underlying number density of all detections, corresponding to the 50th, 80th and 95th percentiles.
    From left to right, the panels display: redshift (with a conversion to fibre diameter in kpc), inclination {for disc galaxies} and stellar mass ($\log M_*$).
    Vertical dashed lines indicate the thresholds for distinguishing high and low velocities at $\pm 50$~\kms.
    In the central inclination panel, blue contours outline the region of parameter space occupied by the systemic component.
}
    \label{fig:hexbin}
\end{figure*}

In \autoref{fig:hexbin}, we map the \ion{Na}{i}~D absorption components in the kinematic plane of Doppler width ($b_D$) versus flow velocity to explore how their host-galaxy properties vary across this space.  
Hexagonal bins show the median value of each property, with individual detections plotted in low-density regions, while contours trace the overall component density. 
{In the middle panel, we only show the inclinations for the subset of disc galaxies.}
Vertical dashed lines mark the $\pm 50$~\kms\ boundaries used to separate low and high velocities and the blue contours in the inclination panel depicts the region associated with the systemic component. 

\subsubsection{Effect of redshift and fibre size}

{We find a pronounced redshift dependence for outflows, with higher-redshift systems exhibiting substantially larger velocities. 
While modest increases in outflow velocity relative to the local Universe are expected at $z \sim 0.3$, the magnitude of the observed trend is unlikely to be driven by evolution alone. 
Instead, the dominant effect is likely a selection bias in the underlying galaxy population: lower-mass galaxies are preferentially observed at lower redshift, while higher-mass systems dominate at higher redshift. 
Given that outflow velocity is known to correlate with stellar mass \citep[e.g.][]{Martin2005, Weiner2009, ChenSDSS2010, Roberts-Borsani2019}, this shift in the mass distribution naturally produces an apparent increase in outflow velocity with redshift. 
In \autoref{fig:hexbin}, regions with low outflow velocities at low redshift correspond to lower stellar masses, as shown in the rightmost panel.}

{
Aperture effects may further contribute to this trend. 
At low redshift, the fibre probes only the central regions of galaxies, whereas at higher redshift it encompasses a larger fraction of the galaxy, increasing the likelihood of intersecting regions where outflows are launched. 
Inflowing absorbers also show a tendency towards higher velocities at increasing redshift, which may reflect a similar combination of selection and aperture effects. 
However, the structure of inflowing gas is less well constrained and disentangling these effects from genuine evolutionary trends remains more uncertain in this case.}

\subsubsection{Inclination}
The influence of galaxy inclination on the detected absorption is evident in the middle panel of \autoref{fig:hexbin}. 
Outflows are generally associated with lower inclinations, with increasing inclination corresponding to smaller velocity offsets until reaching the region dominated by systemic absorption. 
There is tentative evidence that broad inflowing gas occurs at {lower} inclinations compared to the systemic population, although this is limited by the available statistics in this particular figure.

\subsubsection{Stellar mass}
{
We have already noted that the majority of galaxies exhibiting down-the-barrel absorption have stellar masses of $\log M_*/\rm{M_\odot} > 10$. 
In the rightmost panel of \autoref{fig:hexbin}, we illustrate how these stellar mass properties correlate with the measured absorption properties. 
For outflowing gas, higher-velocity winds tend to be launched by galaxies with larger stellar masses, a trend consistent with that reported by \citet{ChenSDSS2010}. 
This dependence likely drives much of the apparent redshift trend discussed in the previous subsection, given that lower-mass galaxies are preferentially observed at low redshift in our sample. }

\subsection{Slow but steady: inflowing gas near systemic velocity}
\begin{figure}
\centering
\includegraphics[width=\linewidth]{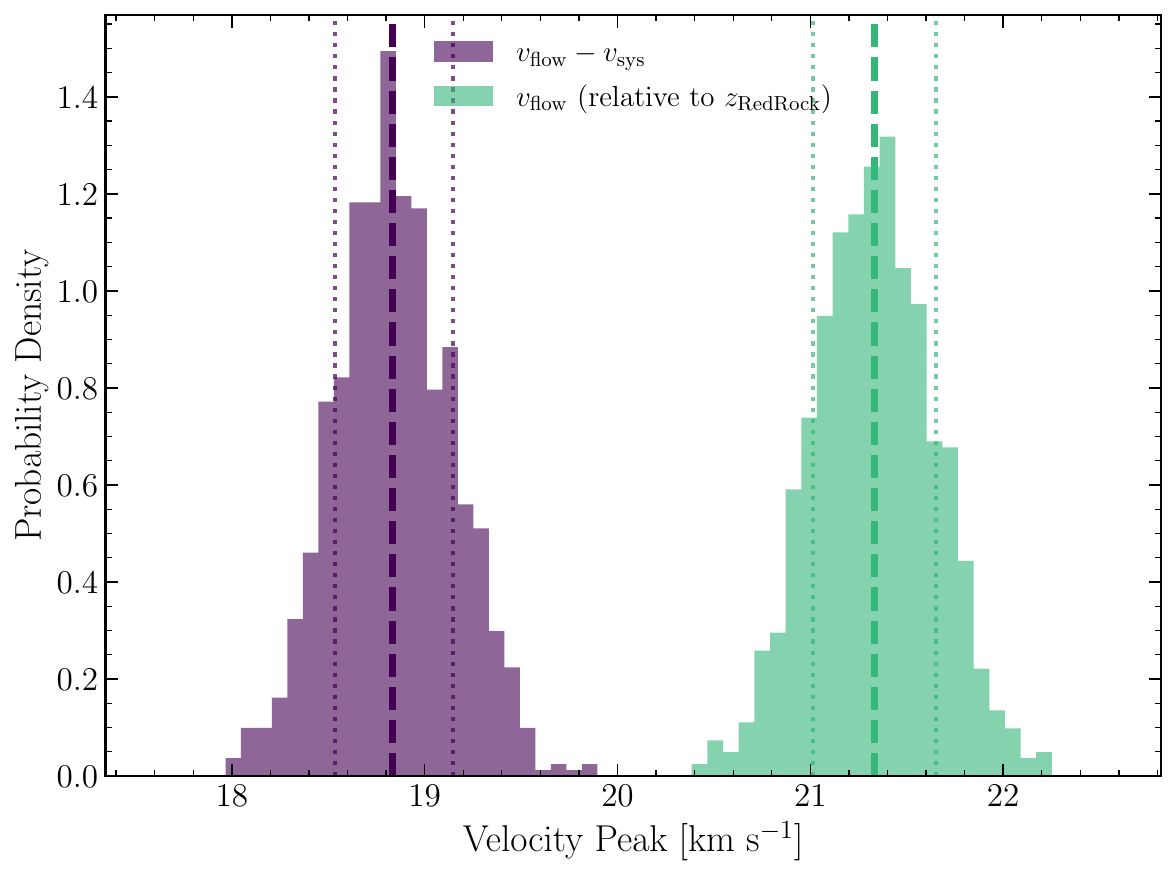}
\caption{
        The distribution of peak gas flow velocities obtained from 1000 Monte Carlo realisations of the velocity distribution.  
        For each realisation we draw from the nested sampling uncertainties and apply a Kernel Density Estimator to determine the location of the peak.  
        The purple histogram shows peak velocities measured relative to the updated systemic velocity derived from \ion{Na}{i}~D, while the green histogram shows peak velocities measured relative to the original \texttt{RedRock} redshift.  
        Dashed vertical lines mark the median of each distribution and dotted lines indicate the 16th and 84th percentiles.
    }
\label{fig:vdist_pos}
\end{figure}

The distribution of gas flow velocities (inset of the bottom left panel of \autoref{fig:param_dist}) shows a clear peak offset from 0~\kms, rather than being centred on zero as would be expected if redshift errors dominated.
To quantify this, we generate 1000 Monte Carlo realisations, perturbing each measurement by its nested sampling uncertainty and estimating the mode of each realisation using a Gaussian KDE.
The resulting distribution (\autoref{fig:vdist_pos}) shows median velocity offsets of $18.8 \pm 0.3$~\kms\ relative to the \ion{Na}{i}~D systemic frame and $21.3 \pm 0.4$~\kms\ relative to the \texttt{RedRock} redshift.
The slightly larger offset in the \texttt{RedRock} frame is consistent with the excess at 10--20~\kms\ seen in \autoref{fig:z_err_dist}.
As redshift errors are expected to be symmetric, the observed shift likely reflects slow inflowing gas that subtly biases the systemic absorption profile.
Although these velocity offsets are similar in magnitude to individual measurement uncertainties, they are statistically significant across the full sample.

The velocity distribution is asymmetric, with a pronounced redshifted tail absent on the blueshifted side.
Many of these low-velocity components have $b_D < 70$~\kms, far narrower than the median stellar velocity dispersion ($\sigma_* \sim 185$~\kms) and are favoured by the Bayesian fits as distinct absorbers.
This implies that absorption within $\pm 50$~\kms of $v_{\rm sys}$ primarily traces low velocity infalling gas rather than the broader systemic component.

Detecting such narrow components requires moderate spectral resolution and large statistical samples.
Previous studies hinted at a significant population of inflows \citep{Sato2009, Sun2024}, but limited resolution and sample sizes hindered secure identification \citep{Carr2022}.
Stacking analyses have produced mixed results, with tentative detections in some cases \citep{Roberts-Borsani2019} and little in others \citep{ChenSDSS2010, Concas2019}.
In our sample, roughly 50\% of inflows with $v_{\rm flow} > 50$~\kms fall in the narrow range 50--100~\kms, increasing to 65\% for a lower threshold of 15~\kms.
Such low velocity features are easily blurred by coarser instrumental profiles and redshift misalignments in stacks.
The presence of inflows in massive galaxies ($\log M_*/\rm{M_\odot} \gtrsim 10.5$) \citep{Roberts-Borsani2019} aligns with our findings, where inflows with $v_{\rm flow} > 50$~\kms have a median stellar mass of $\log M_*/\rm{M_\odot} = 11.2$.
Unlike stacked analyses, the large sample of individual DESI spectra preserves the narrow kinematic structure of these components, enabling detection of inflows previously inaccessible.

There is also a smaller but significant population of narrow outflows with $-50 < v_{\rm flow} < 0$~\kms.
While the overall outflow distribution peaks near {$-$250~\kms (\autoref{fig:param_dist})}, these slow outflows likely reflect geometric effects, being more common in higher inclination galaxies where the line of sight is close to perpendicular to the wind direction (\autoref{fig:hexbin}).
Together, these populations highlight DESI's ability to detect and characterise gas flows that were unresolved or blended in previous spectroscopic surveys.

\subsection{On the origins of inflows}
{
Galaxies require a continuous supply of gas to sustain their star-formation rates \citep{Saintonge2017, Tacconi2018}.
In the context of galaxy formation, gas accretion is often described in terms of hot-mode and cold-mode accretion, where the latter is expected to dominate in lower-mass haloes and at higher redshift, allowing gas to accrete without being shock-heated to the virial temperature \citep[e.g.][]{BirnboimDekel2003, Dekel2006}. 
{At lower redshift and in more massive systems, gas is instead thought to accrete primarily via the cooling of a hot circumgalactic medium \citep{Maller2004, Faucher2011, vandeVoort2011, Voit2015, Stern2020, Stern2021, Trapp2022, Trapp2024}, supplying less-processed material from large ($\sim$10--50 kpc) scales.}
On smaller spatial scales (a few kpc), galactic fountains recycle enriched material between the disc and lower halo. 
In the Milky Way, high- and intermediate-velocity clouds provide evidence that a substantial fraction of the cool gas reaching the disc is linked to this process \citep{Putman2012, Marasco2022}, while neutral extraplanar gas observed around nearby \ion{H}{i} discs is also broadly consistent with a fountain origin \citep{Fraternali2017, Li2023}. 
Such recycling can dominate the inferred accretion rate in some simulations \citep{Lucchini2024, Barbani2025}.
These processes are not mutually exclusive and may operate simultaneously, with gas cycling between the disc and halo while also being supplied from larger scales. 
However, it remains debated which mechanism dominates the net inflow of gas onto galaxies.}

In IFS observations, \citet{Roy2021, Moghni2026} find patchy inflow signatures across the discs of early-type ‘red geyser’ galaxies, with mass inflow rates consistent with minor mergers or fountain-driven accretion.
\citet{Rupke2021} report that, in eight nearby AGN-host galaxies, inflows are preferentially detected along the projected major axis and may arise from non-axisymmetric potentials, tidal motions, or halo infall.
Earlier work also provides insight into how infall may occur.
\citet{Rubin2012} detect inflows primarily in galaxies with inclinations $\gtrsim 55^\circ$, with velocities of 80–200 \kms.
\citet{Martin2012} identify inflows in four out of nine galaxies where an additional object appears in the slit at a similar redshift, suggesting extended structures intersecting the line of sight.
More recently, \citet{Sun2024} find that positive velocity offsets (tracing infalling gas) are preferentially found in quiescent galaxies. 
\citet{Antonia2025}, combining ultra-strong \ion{Mg}{ii} absorption in QSO sightlines with down-the-barrel galaxy spectra, find five inflows consistent with fountain-driven, filamentary and minor-merger-driven accretion.
Although some of these studies rely on \ion{Mg}{ii}, which traces more ionised gas and therefore structures potentially located further from the disc than \ion{Na}{i} D \citep{Rubin2012, Martin2012, Antonia2025}, they remain highly informative.
A wide range of other works have also identified inflows \citep{Floch2007, Sato2009, Krug2010, Coil2011, Sarzi2016, Rubin2017Review, Johnson2022, Weldon2023, Coleman2024, Bevacqua2026}.

Together, these simulations and observations indicate that gas accretion, while essential to galaxy evolution, may not necessarily proceed through a single pathway.
The kinematic properties of inflows in \autoref{fig:hexbin} span a broad range of velocities and line widths, implying multiple accretion channels whose relative importance likely depends on galaxy properties.
Therefore, this large population of $\sim$10 000 inflows in this work likely traces a mixture of physical phenomena.

Importantly, not all kinematically identified inflows or outflows correspond to genuine accretion or feedback events (as mentioned first in \autoref{sec:results}).
Some absorbers are ``false friends'', in the sense that they exhibit inflow- or outflow-like velocities along the line of sight but arise from unrelated phenomena such as late-stage mergers.
These cases mimic the kinematic signatures of gas flows without directly tracing the baryon cycle of the host galaxy itself. 
We discuss these various cases together in the following sections. 

\subsubsection{Bars and radial inflows}
{A significant population of down-the-barrel absorbers lie at velocities $v_{\rm flow} < 100$ \kms with small Doppler parameters $b_D < 50$ \kms and are preferentially found in more edge-on galaxies (middle panel of \autoref{fig:hexbin}).
These properties: low velocities, narrow line widths and high inclinations, are broadly consistent with coherent, planar gas motions within the disc.
One natural explanation is bar-driven inflow: bars are elongated, non-axisymmetric stellar structures capable of funnelling gas toward the central regions of galaxies \citep{deV1963, Miller1970}.
Typical bar-driven inflow velocities are $\lesssim 100$ \kms with relatively small dispersions, in good agreement with the observed kinematics.
This phenomenon may therefore explain a subset of the slow, low-dispersion infall we observe. }
However, the relatively high covering fractions of our detections ($C_f \sim 0.5$) imply that these features are only measurable when the fibre subtends a physical scale comparable to the bar. 
Consequently, bars are likely to contribute only to the inflows detected at $z \lesssim 0.15$ where the fibre diameter extends 4 kpc. 
Furthermore, only a minority of our inflows can be attributed to bars, as most detections occur in early-type galaxies (\autoref{fig:morph}).
For similar reasons, the tidal streaming seen in spiral arms \citep{Shetty2007} likely accounts for only a fraction of the observed inflows detected here. 

An alternative is that the inflows we observe represent the final stage of circumgalactic gas accreting onto the disc. 
\ion{Mg}{ii} absorption towards QSOs reveals a co-rotating component extending to large radii in the CGM \citep{Ho2017, Zabl2019}.
By contrast, \ion{Na}{i} D traces colder, dust-rich gas and therefore likely probes material closer to the disc, as it is unclear if the low-redshift CGM contains sufficient dust for \ion{Na}{i} to survive \citep{Menard2010, Peek2015}.
Nonetheless, the typical infall velocities of $\lesssim 60$ \kms reported by \citet{Ho2019} are consistent with the bulk of our inflowing population, for which half of the absorbers with $v_{\rm flow} > 0$ \kms have velocities below 60 \kms.
We also find that the velocity distribution peaks at approximately 20 \kms (\autoref{fig:vdist_pos}), indicating a substantial population of slow, infalling gas.
{This is in line with the prediction of \citet{Trapp2024}, who find a time-averaged radial velocity of $\sim$20 \kms at $\sim$2$R_{\rm DLA}$ (30--40 kpc), where $R_{\rm DLA}$ marks the radius at which the neutral hydrogen column density drops below the damped Ly$\alpha$ (DLA) threshold, $\log N_{\rm HI} = 20.3$.}
Although we do not estimate \ion{H}{i} column densities for the inflows in this work, they often have DLA-like columns in the literature \citep[e.g.][]{Krug2010}.
Given the uncertainties in ionisation and dust-depletion corrections, it remains plausible that \ion{Na}{i} D detected further from the disc traces somewhat lower \ion{H}{i} densities, owing to reduced dust content and higher ionisation. 
These low-velocity streams would then represent the final stages of gas accretion, settling onto the outer disc before migrating inward. 
If the inflowing gas is warped with respect to the stellar disc \citep{Sankar2025}, this geometry could naturally explain the median inclinations of $\sim$63$^\circ$ observed for galaxies hosting inflowing absorbers.
In this picture, warped inflows are difficult to detect in perfectly edge-on systems due to low covering fractions, but become more readily observable at slightly smaller inclinations where the line of sight intersects the accreting material. 

The majority of our inflow detections occur in early-type galaxies, consistent with previous measurements \citep{Sun2024}.
Although ETGs typically have lower star-formation rates than late-type galaxies, they are not devoid of cold gas: many host molecular or \ion{H}{i} reservoirs, albeit with a lower incidence than in spirals \citep[e.g.][]{vanDriel1991, Bregman1992, Morganti2006, Young2011, Serra2012}.
In particular, \ion{H}{i} discs in ETGs can extend well beyond the stellar body, reaching tens of kpc \citep{Serra2014} and thus, provide a substantial reservoir from which gas can gradually lose angular momentum and migrate inwards. 
The low velocities and small Doppler parameters of our down-the-barrel absorbers are therefore consistent with tracing the slow inward transport of gas from an extended \ion{H}{i} disc into the central stellar component.
Such a supply of gas may help sustain the young stellar populations observed in the cores of ETGs \citep{Trager2000, Kuntschner2010}.

\subsubsection{Fountains}
\begin{figure*}
    \centering
    \includegraphics[width=\linewidth]{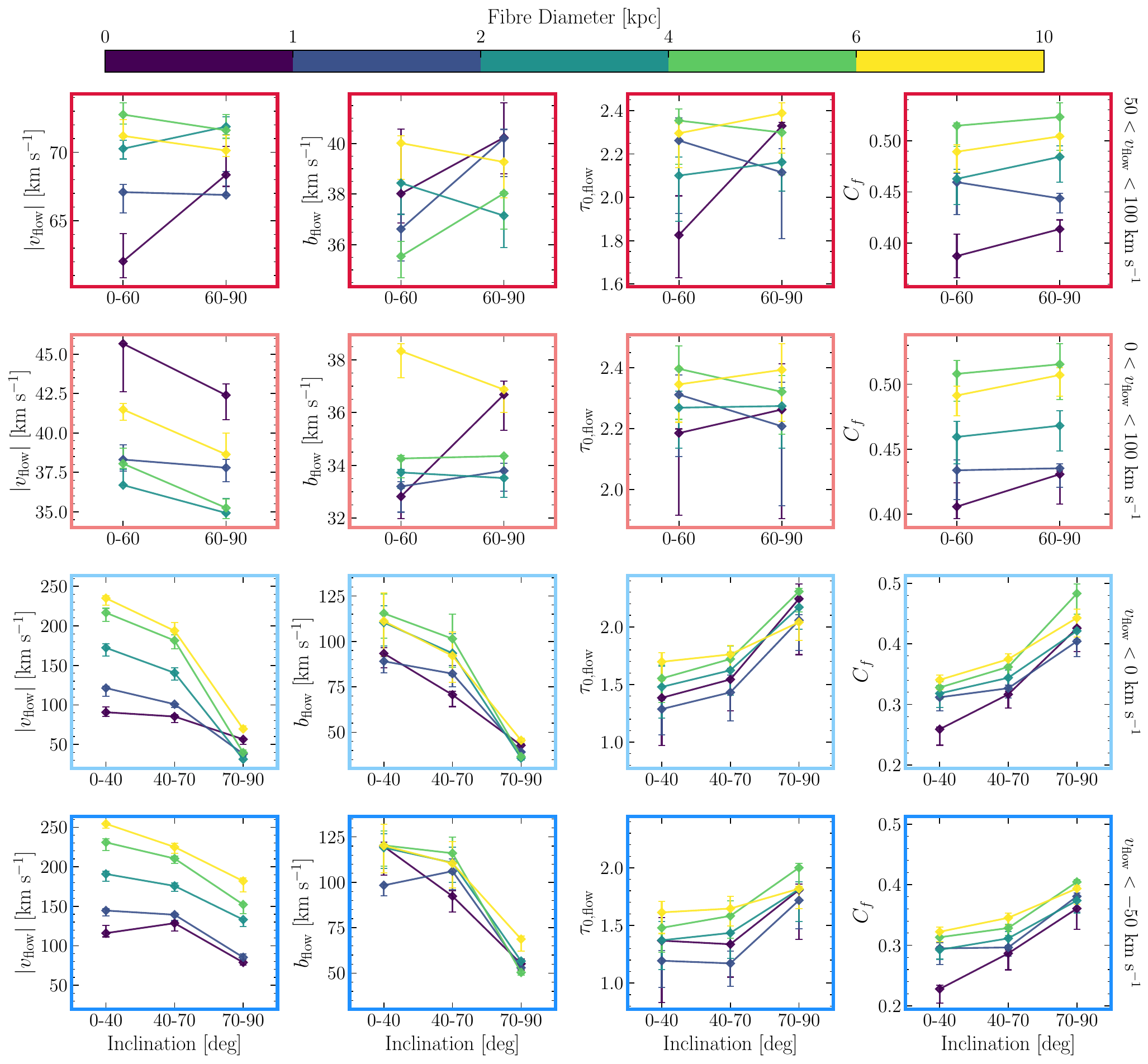}
    \caption{
        Variation of gas-flow properties {in disc galaxies} with inclination at fixed physical diameter subtended by the DESI fibre. 
        From left to right, we show the median velocity offset from systemic, Doppler parameter, optical depth at line centre for the D1 line and covering fraction. 
        {The middle two rows include the inflow and outflow samples shown in the top and bottom rows, respectively; the row labels indicate the velocity range used to define each plotted subsample.}
        Top to bottom, the panels correspond to inflows, redshifted components, blueshifted components and outflows. 
        We use two or three inclination bins with values [0, 60), (60, 90] and [0, 40), (40, 70] and (70, 90] degrees, respectively for inflows and outflows. 
        Coloured lines indicate different effective fibre diameters in kpc, as shown in the top colour bar. 
        Only disc galaxies are included here, corresponding to the sample shown in the middle panel of \autoref{fig:hexbin}. 
        }
    \label{fig:fix_inc_z}
    \end{figure*}

Galactic fountains have been proposed as a dominant channel for gas recycling within disc galaxies \citep{Fraternali2017, Marasco2022, Lochhaas2025}. 
A key prediction of this scenario is that recycled gas returns to the disc from above, rather than being accreted predominantly at its edges.
In \autoref{fig:fix_inc_z}, we examine how the median measured gas properties vary with {the inclination of disc galaxies}, separating the sample by the physical diameter probed by the fibre.  
We show these trends for inflows with $v_{\rm flow} > 50$~\kms, redshifted absorbers with $v_{\rm flow} > 0$~\kms, blueshifted absorbers with $v_{\rm flow} < 0$~\kms and outflows with $v_{\rm flow} < -50$~\kms from top to bottom. 
We additionally require inflows to have $v_{\rm flow} < 100$~\kms to minimise contamination from late-stage mergers and satellite accretion (refer to the forthcoming section).
Each line in the plots corresponds to a different fibre diameter, as indicated by the colour bar and represents the physical scale covered in kpc.  
The points show the median value in bins of inclination: [0, 60) and [60, 90] for inflows and [0, 40), (40, 70] and (70, 90] degrees for outflows.  
Error bars indicate the 16th to 84th percentiles of the distribution of medians derived from 1000 Monte Carlo realisations of the sample after accounting for measurement uncertainties.

{If inflows arise from galactic fountains, the returning gas may preferentially have a velocity component perpendicular to the disc. 
However, fountain-driven accretion is not expected to be purely vertical: interactions with the surrounding halo may produce substantial radial and rotational motions \citep{Fraternali2008}. 
Its projected velocity may therefore exhibit no straightforward dependence on inclination. 
Indeed, we find little evidence for such a trend, with differences in inflow velocity consistent within $1\sigma$ for almost all fibre diameters. 
Any inclination dependence is consequently weak, in contrast to the clearer behaviour observed for outflows. 
We also find no clear signature of predominantly coplanar radial inflow, although such a trend could be obscured by variations in intrinsic inflow speed and the orientations of bars or other non-axisymmetric structures. 
The absence of a clear inclination dependence therefore provides only limited discrimination among these scenarios.}

{A substantial fraction of our detections occur in galaxies classified as early types, where fountain-driven accretion may be less prevalent because it requires ongoing star formation to eject gas from a disc. 
However, this morphological classification does not preclude the presence of a disc or ongoing star formation, particularly at higher redshift. 
Alternatively, cosmological accretion may produce cool inflowing clouds in the haloes of early-type galaxies, although such clouds may be disrupted through interactions with the hot halo before reaching the central galaxy \citep{Afruni2019}. 
Our sensitivity to fountains may also be limited: we are highly incomplete at velocities $\lesssim 100$~\kms and dispersions $b_D \gtrsim 70$~\kms. 
In the Milky Way and similarly sized nearby galaxies, typical outflow and inflow speeds are of order 50--100~\kms \citep{Marasco2022, Li2023}. 
Slowly inflowing gas, particularly when distributed over a broad range of velocities, may be difficult to distinguish from systemic absorption and could therefore be missed by our current methods.}

\subsubsection{Mergers and infalling satellites}
{
High-velocity inflows of cool gas are not generally expected in standard models of smooth gas accretion at $z < 0.6$.
Cooling-flow accretion and co-rotating CGM structures typically produce radial velocities of $\lesssim 60$ \kms \citep{Ho2019, Trapp2022}, while even at higher redshift inflowing gas in the disc rarely exceeds $\sim$100 \kms \citep{Genzel2023, Jolly2026}.
Similarly, fountain-driven recycling is limited by the initial ejection velocity and is therefore unlikely to reach $> 100$ \kms observed in our highest-velocity systems. 
Although gas freely falling very close to a supermassive black hole can attain higher speeds \citep{vangorkom1989, Yoon2025FLASH}, most of our high-velocity absorbers occur at higher redshift, where the DESI fibre subtends a larger physical scale and probes beyond the nuclear regions.
The presence of such fast inflows in \autoref{fig:hexbin} therefore requires an alternative origin.}

\begin{figure}
    \centering
    \includegraphics[width=\linewidth]{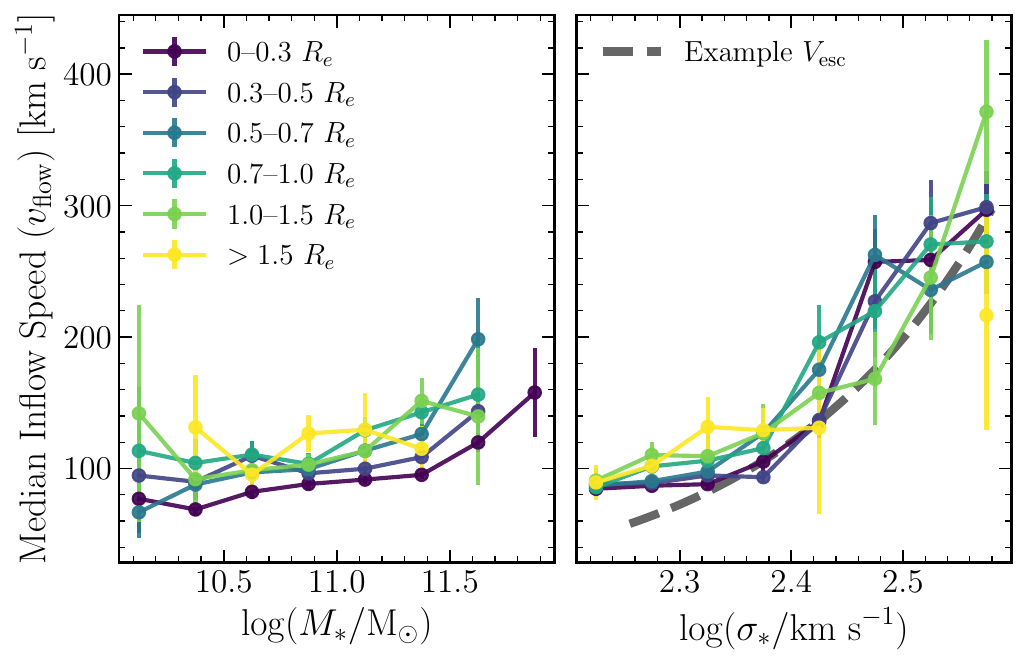}
    \caption{
    Median inflow velocity as a function of stellar mass (left) and stellar velocity dispersion (right) for early-type galaxies.
    Colours indicate the fraction of the effective radius covered by the DESI fibre.
    In the right panel, the dashed line shows a representative escape-velocity curve, approximating the maximum expected infall speed.
    }
    \label{fig:ETGInflow}
\end{figure}

To investigate the origin of this fast infalling gas, we show in \autoref{fig:ETGInflow} the median inflow velocity for early-type galaxies as a function of stellar mass (left) and stellar velocity dispersion (right). 
The correlation with velocity dispersion is significantly stronger than with stellar mass, suggesting that the inflow speed is primarily governed by the depth of the gravitational potential. 
For reference, the dashed line in the right panel shows a representative escape-velocity curve, comparable to the maximum expected infall speed. 
The trend is consistent across different fibre coverages (normalised by the effective radius), indicating that it is not driven by aperture effects. 
The plateau near $\log\sigma_* \sim 2.2$ is partially driven by the velocity limit of $50$~\kms used to define inflows. 

{
Given these constraints, radial flows and fountain-driven recycling are unlikely to account for the highest-velocity inflows in our sample.
In contrast, infalling satellites or merging systems can naturally reach near-ballistic velocities within the host potential, providing a straightforward explanation for cool gas observed at several hundred \kms. 
Indeed, in the nine inflow cases reported by \citet{Martin2012}, four show emission-line components potentially associated with a second galaxy at the inflow velocity and the fastest inflow ($\sim$400~\kms) is accompanied by an emission component offset by $\sim$350~\kms. 
The strong correlation between inflow velocity and stellar velocity dispersion (\autoref{fig:ETGInflow}) further supports this picture.
We therefore conclude that a significant fraction of the highest-velocity inflows ($v_{\rm flow} > 100$ \kms) are likely associated with mergers or satellite accretion, rather than smooth gas inflow or recycling.}

\subsection{On the origins of outflows}
\subsubsection{High-speed narrow outflows}
In the kinematic phase-space diagrams of \autoref{fig:morph} and \autoref{fig:hexbin}, we identify a population of strongly blueshifted absorbers with narrow line widths that is challenging to interpret within the standard picture of galactic-scale outflows. 
While these features are robust detections, their kinematic properties place them at the extreme edge of parameter space explored by both simulations and most observational studies. 
As a result, their physical origin is not immediately clear.

A natural framework for interpreting these systems is provided by wind-blown bubble models \citep{Xu2022}, in which the momentum flux of a wind drives an expanding shell into the ISM and CGM. 
In this geometry, absorption along the line of sight samples only a limited portion of the near side of the bubble directly in front of the galaxy, leading to narrow blueshifted profiles with $b_D \ll v_{\rm flow}$. 
This provides a simple geometric explanation for producing high-velocity, narrow absorption without requiring highly collimated outflows. 
While \citet{Xu2022} do not report similarly extreme narrow components, their analysis focuses on starburst galaxies and primarily traces warmer gas phases, which may exhibit broader kinematics than the colder material probed by \ion{Na}{i}~D. 
If the observed shell structures are long-lived remnants of past star formation activity, the resulting absorption need not be directly linked to ongoing or recent star formation.

Narrow, high-velocity outflows have been observed in compact starburst and post-starburst galaxies, with $v_{\rm flow} > 1000$~\kms and $b_D \sim 10$~\kms traced by \ion{Mg}{ii} \citep{Tremonti2007, Diamond2012, Perrotta2023}, consistent with rapidly expanding shells. 
Such examples are also seen in systems hosting radio jets, such as Mrk~6, where \ion{Na}{i}~D absorption with $b_D = 16$~\kms is observed at velocities below $-1000$~\kms \citep{Krug2010}. 
While jet--ISM interactions are expected to broaden the overall outflow, localised regions of compressed gas may give rise to narrow absorption components on sub-kiloparsec scales \citep{Girdhar2022, Ward2024}. 

At the same time, alternative explanations unrelated to outflows from the target galaxy remain plausible. 
In particular, narrow high-velocity components could arise if the observed continuum is intersected by an unrelated foreground gas along the line of sight. 
In this scenario, absorption from gas that is not gravitationally bound to the target galaxy and participating in the Hubble flow could imprint narrow features at negative apparent velocity offsets, mimicking narrow down-the-barrel outflows. 
However, such an interpretation would be subject to constraints from the incidence rates of intervening absorbers ($dN/dz$), which remain poorly constrained for \ion{Na}{i}~D. 
Existing observations demonstrate that \ion{Na}{i}~D absorption can arise in foreground systems along quasar sightlines \citep{Carilli1992, Weng2022} and that metal-line absorption such as \ion{Mg}{ii} is commonly detected towards background galaxies \citep{Peroux2018, Augustin2021}. 
While this indicates that intervening absorption is a viable contaminant, its overall contribution to the population of narrow, high-velocity features in our sample is difficult to quantify in the absence of robust $dN/dz$ measurements for \ion{Na}{i}~D.

Taken together, these considerations suggest that no single mechanism is likely to explain all of the narrow, highly blueshifted absorbers in our sample. 
A combination of wind-blown bubbles, radio jets and line-of-sight contamination may all contribute.

\begin{figure}
    \centering
    \includegraphics[width=\linewidth]{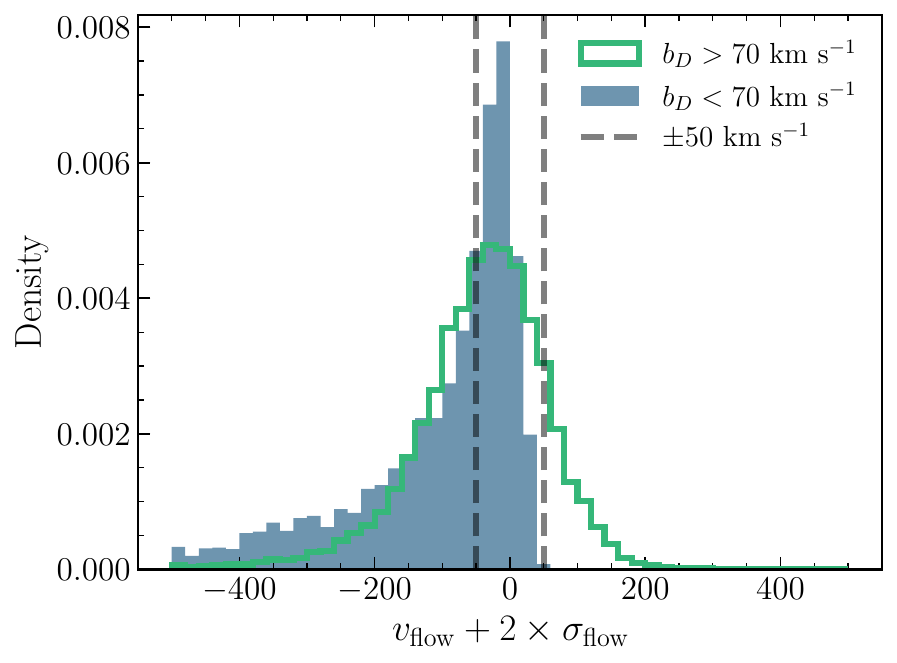}
    \caption{
    The distribution of net outflow blueshift relative to the line width for narrow ($b_D < 70$ \kms) and broad ($b_D \ge 70$ \kms) outflows, following \citet{Heckman2000}. 
    Here, $v_{\rm flow} + 2\sigma_{\rm flow}$ is shown, where $\sigma = \sqrt{2} b_D$. 
    }
    \label{fig:broad_narrow_out}
\end{figure}

To further investigate the origin of this population of narrow absorbers, we examine the distribution of net blueshift relative to the line width in \autoref{fig:broad_narrow_out}, separating the sample into narrow and broad components using $b_D = 70$~\kms\ as the dividing threshold \citep{Heckman2000}.
We show $v_{\rm flow} + 2\sigma_{\rm flow}$, where $\sigma_{\rm flow} = \sqrt{2}b_D$, which captures the velocity extent of the absorption on the redshifted side.

Broad outflows are approximately symmetric about zero, indicating that the reddest absorption occurs close to the systemic velocity. This behaviour is consistent with gas that is initially at rest and subsequently accelerated outward.
Narrow absorbers also peak near zero, suggesting that many are consistent with genuine outflows; however, they exhibit a pronounced tail towards negative velocities.
This tail indicates that a subset of these absorbers may arise from cases where the apparent blueshift is dominated by the Hubble flow between background and foreground galaxies.

Taken together, these considerations suggest that the population of high-speed, narrow absorbers likely arises from a mixture of physical origins. These include genuinely extreme outflows in rare systems, as well as apparent high-velocity components produced by galaxy–galaxy alignments.
Disentangling these scenarios will require constraints on AGN properties and galaxy environments. 

\subsubsection{The extent and geometry of outflows}
While outflows are generally better studied than inflows, there are nonetheless interesting trends. 
We find that both the outflow speed and line width decrease with increasing inclination for all fibre diameters, consistent with previous studies \citep{Weiner2009, ChenSDSS2010, Bordoloi2014, Rubin2014, Roberts-Borsani2019, Sun2024}.
{The roughly constant outflow velocity in the bottom two inclination bins of \autoref{fig:fix_inc_z}, followed by the sharp decrease in the highest-inclination bin, is consistent with a biconical wind geometry similar to that of M82, with an approximate opening angle of 70$^\circ$.
However, this estimate is sensitive to the adopted inclination binning and should therefore be interpreted as approximate.}

There are also notable differences compared to previous results.
\autoref{fig:hexbin} reveals a significant population of narrow ($b_D \lesssim 70$~\kms) outflows that has not been reported in earlier surveys. 
These systems are predominantly associated with {high-inclination} galaxies and exhibit relatively low velocities. 
In \autoref{fig:fix_inc_z}, the line widths decline sharply between the second and third inclination bins, reaching an approximately constant value of 40--60~\kms across all fibre diameters and redshifts, highlighting this previously unresolved population.
This behaviour may indicate that gas near the base of the outflow is kinematically cold and slow-moving. 
Alternatively, some of these signatures could arise from disc instabilities or other internal dynamical processes that produce narrow absorption components close to the systemic velocity. 
Such components are likely suppressed in earlier work owing to limited spectral resolution and the tendency of stacking analyses to wash out narrow absorption near the systemic velocity.

\subsection{On the incidence and covering fraction of outflows and inflows}
The line-of-sight covering fraction, $C_f$, appears to be systematically higher for inflows than for outflows. 
We emphasise that this per-sightline quantity, derived from the absorption profile, is distinct from the global covering fraction or incidence, which measures the fraction of galaxies (or sightlines) showing detectable absorption within a given velocity range.
Despite the larger line-of-sight $C_f$ for inflows, the incidence of outflows remains more than twice that of inflows in our sample. 
This is consistent with previous signal-to-noise limited \ion{Na}{i} D surveys, such as \citet{Sato2009}. 
A comparable outflow fraction {(seen in \autoref{tab:velocity_stats})} has been reported for starburst and post-starburst galaxies in SDSS, with a higher fraction of inflows in quiescent systems \citep{Sun2024}. 

{
In contrast, the incidence of \ion{Mg}{ii} inflows appears to be substantially lower \citep{Martin2012,Rubin2012}. 
However, this comparison is subject to important selection effects.
\ion{Mg}{ii} studies are typically conducted at higher redshift in order to shift the doublet into the optical window, resulting in smaller samples that are often biased towards galaxies with higher star-formation rates. 
In addition, the lower spectral resolution typically achieved at bluer wavelengths makes it more difficult to resolve the narrow inflowing absorbers identified in this work. 
Finally, \ion{Mg}{ii} traces warmer, more ionised gas than \ion{Na}{i} and may probe a different phase that is less commonly detected in low-dispersion, coherent inflowing structures.}

The higher incidence of outflows may imply that their global covering fraction is larger than that of inflows. 
However, if the relative incidence of inflows and outflows depends on galaxy type, as shown by \citet{Sun2024}, this difference may instead reflect the underlying population of galaxies in our sample. 
Ongoing work is exploring this using a more complete set of galaxy properties, particularly the star-formation rate.
Nevertheless, the higher line-of-sight covering fraction for inflows is intriguing. 
At face value, it suggests that the projected area of inflowing gas across the stellar continuum is larger than that of outflows. 
An alternative explanation is that inflows are preferentially located near the galaxy centre, where the surface brightness is highest, leading to a larger light-weighted covering fraction. 
Integral field spectroscopy will be essential for distinguishing between these possibilities.

\subsection{Future methodological improvements}
While the current pipeline is robust, several improvements are planned for future DESI releases.

\begin{itemize}
    \item We currently model stellar and systemic interstellar absorption as a single component, which limits our ability to recover the kinematics of gas near systemic velocity. 
    Implementing an expanded set of stellar templates or higher-resolution (in parameter space) spectral libraries may allow the stellar \ion{Na}{i} D absorption to be more precisely constrained, enabling the measurement of bulk motions within the systemic gas and the detection of emission. 

    \item Our nested model framework assumes that a systemic component is always present. 
    In cases where it is absent, this introduces unnecessary prior volume and reduces completeness (\autoref{fig:completeness_1D}). 
    A dedicated selection pathway for flow-only models would therefore improve the completeness.

    \item {Our use of broad uniform priors is conservative and makes it harder for gas flow models to be favoured by the Bayesian evidence. }
    Adopting more physically motivated priors could further improve completeness, although defining these in an unbiased way remains challenging.
    
    \item Some high-SNR spectra exhibit clear \ion{Na}{i}~D emission or require additional flow components. 
    Extending the model to include emission and more complex absorption structures would improve fits in these systems.
    
    \item The computational cost of nested sampling limits scalability and extensive testing. 
    Simulation-based inference offers a promising alternative, enabling orders-of-magnitude faster posterior estimation and model comparison \citep{Mancini2023, Aufort2025}.
    \end{itemize}

\subsection{Future work}
This work has primarily focused on the methodologies developed to identify and characterise down-the-barrel gas flows, alongside an initial investigation into how stellar and orientation properties inform the origins of this gas. 
Given the unprecedented scale and quality of the DESI dataset, several compelling avenues for future research remain to be explored.

There remains uncertainty about the origins of gas accretion, as the down-the-barrel absorption we measure could arise from any location within the fibre aperture between the galaxy and observer. 
Future work with integral-field spectrographs would allow spatially resolved mapping of the gas \citep{Roy2021, Rupke2021, Moghni2026} to better understand its origin. 

We have so far explored trends with basic galaxy properties. 
Extending this to star formation rate, metallicity, AGN activity and environment will clarify the drivers of inflows and outflows. 
Measuring mass inflow and outflow rates is also essential to assess whether galaxies can sustain their star formation or are approaching quenching.

Higher-SNR subsamples allow reconstruction of star-formation histories using tools such as \texttt{PPXF}, enabling direct tests of the temporal connection between gas flows and stellar assembly \citep{Belli2024}.
Stacking analyses will extend sensitivity to lower-mass galaxies, which are currently not probed in this work.
We also aim to connect the interstellar and circumgalactic media by combining these measurements with background QSO sightlines probing galaxy haloes \citep{Kacprzak2014, Rahmani2018, Peroux2019, Peroux2022, Weng2023, Weng2023a, Antonia2025}.

With the upcoming internal release of millions more spectra in DESI DR3 and beyond, we anticipate that the sample of down-the-barrel detections will expand significantly over the coming years. 
Complementary surveys such as WEAVE \citep{Jin2024} and 4MOST \citep{Jong4MOST2012} offer higher spectral resolution, which will be instrumental in resolving kinematic components at low velocity offsets that may be blended at the DESI resolution. 
Concurrently, facilities including the Subaru Prime Focus Spectrograph (PFS; \citealp{PFS}) and VLT/MOONS \citep{Moons} will enable us to probe the \ion{Na}{i} D doublet at higher redshifts in large samples. 
Together, these surveys will facilitate an analysis on how galactic outflows and inflows have evolved over cosmic time, providing a more complete view of the baryon cycle across different epochs.

\section{Conclusions}
\label{sec:conclusion}
We present the largest sample of down-the-barrel \ion{Na}{i}~D absorbers to date, derived from the DESI Data Release 2. 
We use nested sampling to compute evidences and Bayes factors, to determine whether complex models involving gas flows are statistically required over simpler models. 
Our main conclusions are summarised as follows:

\begin{itemize}
    \item {Sample statistics:} 
    From an initial parent sample of $\sim$6 million galaxies with signal-to-noise ratio $> 5$, we identify 50 088 galaxies exhibiting statistically significant down-the-barrel absorption ($\Delta \ln \mathcal{Z} > 1.0$). 
    This sample comprises 50 835 individual absorption components. 
    Using a velocity threshold of $\pm 50$ \kms\ relative to the systemic redshift, we classify 48.7\% of these components as outflows, 22.3\% as inflows and 29.0\% as systemic absorption. 
    This catalogue represents an increase of almost two orders of magnitude compared to all previous studies.

    \item {A large population of slow inflows:} 
    We measure the properties of a significant population of absorbers tracing infalling gas for the first time. 
    These components are characterised by narrow line widths, typically $b_D < 70$ \kms and high optical depths. 
    The spectral resolution of DESI allows us to resolve these features, which were likely washed out in previous low-resolution studies or stacking experiments.
    We find evidence that much of the absorption classified as `low-velocity' ($|v| \le 50$ \kms) actually traces slow-infalling gas rather than bulk ISM motion. 
    The peak of the velocity distribution for these components is offset by $\sim +20$ \kms\ relative to the systemic stellar redshift. 
    Furthermore, their Doppler widths are significantly narrower than the stellar velocity dispersion, ruling out the rotating disc as their primary origin. 

    \item {Evidence for various accretion modes:}
    Our observations suggest that multiple mechanisms contribute to gas accretion onto galaxies. 
    Slow, low-dispersion inflows may arise from radial motions in disc galaxies or from the gradual settling of circumgalactic gas, whereas higher-velocity inflows are consistent with accretion from satellites or minor mergers. 
    Galactic fountains may also contribute. 
    Although we find no significant dependence of inflow velocity on inclination, fountain-driven accretion can involve a combination of vertical, radial and rotational motions and may therefore lack a simple inclination signature. 
    The interpretation is further limited by our sensitivity to slow, broad components. 
    Integral-field spectroscopy will be crucial for spatially resolving these flows and distinguishing among the possible accretion channels. 
    Overall, our results indicate a complex picture in which several modes of accretion may operate, with their relative importance influenced by galaxy morphology and environment.
    
    \item {A large population of outflows:} 
    We detect almost 25 000 absorbers consistent with outflows, with velocities $< -50$ \kms relative to the systemic galaxy redshift. 
    Outflow properties show a strong dependence on inclination, with decreasing velocity and line width, and increasing optical depth, as galaxies become more edge-on. 
    In addition, we identify a population of low-speed, narrow outflows preferentially found in edge-on systems. 
    Some of the high-velocity ($< -500$ \kms) narrow outflows may trace shell structures.    
    \end{itemize}

This study demonstrates the power of DESI to advance our understanding of the baryon cycle. 
With a sample of {50 088} down-the-barrel absorbers, we can move beyond stacking analyses to directly reveal the complex nature of gas flows in the low-redshift Universe. 
Future work will use this sample to quantify mass flow rates and examine their connection to star-formation and AGN activity and large-scale structure, offering a comprehensive picture of how galaxies acquire, cycle and return gas to their surroundings.

\begin{acknowledgements}
 This material is based upon work supported by the U.S. Department of Energy (DOE), Office of Science, Office of High-Energy Physics, under Contract No. DE–AC02–05CH11231, and by the National Energy Research Scientific Computing Center, a DOE Office of Science User Facility under the same contract. Additional support for DESI was provided by the U.S. National Science Foundation (NSF), Division of Astronomical Sciences under Contract No. AST-0950945 to the NSF’s National Optical-Infrared Astronomy Research Laboratory; the Science and Technology Facilities Council of the United Kingdom; the Gordon and Betty Moore Foundation; the Heising-Simons Foundation; the French Alternative Energies and Atomic Energy Commission (CEA); the Secretariat of Science, Humanities, Technology and Innovation (SECIHTI) of Mexico; the Ministry of Science, Innovation and Universities of Spain (MICIU/AEI/10.13039/501100011033), and by the DESI Member Institutions: \url{https://www.desi.lbl.gov/collaborating-institutions}.
\newline
The DESI Legacy Imaging Surveys consist of three individual and complementary projects: the Dark Energy Camera Legacy Survey (DECaLS), the Beijing-Arizona Sky Survey (BASS), and the Mayall z-band Legacy Survey (MzLS). DECaLS, BASS and MzLS together include data obtained, respectively, at the Blanco telescope, Cerro Tololo Inter-American Observatory, NSF’s NOIRLab; the Bok telescope, Steward Observatory, University of Arizona; and the Mayall telescope, Kitt Peak National Observatory, NOIRLab. NOIRLab is operated by the Association of Universities for Research in Astronomy (AURA) under a cooperative agreement with the National Science Foundation. Pipeline processing and analyses of the data were supported by NOIRLab and the Lawrence Berkeley National Laboratory. Legacy Surveys also uses data products from the Near-Earth Object Wide-field Infrared Survey Explorer (NEOWISE), a project of the Jet Propulsion Laboratory/California Institute of Technology, funded by the National Aeronautics and Space Administration. Legacy Surveys was supported by: the Director, Office of Science, Office of High Energy Physics of the U.S. Department of Energy; the National Energy Research Scientific Computing Center, a DOE Office of Science User Facility; the U.S. National Science Foundation, Division of Astronomical Sciences; the National Astronomical Observatories of China, the Chinese Academy of Sciences and the Chinese National Natural Science Foundation. LBNL is managed by the Regents of the University of California under contract to the U.S. Department of Energy. The complete acknowledgments can be found at \url{https://www.legacysurvey.org/}.
\newline
Any opinions, findings, and conclusions or recommendations expressed in this material are those of the author(s) and do not necessarily reflect the views of the U. S. National Science Foundation, the U. S. Department of Energy, or any of the listed funding agencies.

The authors are honored to be permitted to conduct scientific research on I'oligam Du'ag (Kitt Peak), a mountain with particular significance to the Tohono O’odham Nation. 

\newline
This work has made use of CosmoHub \citep{TALLADA2020100391, 2017ehep.confE.488C}, developed by PIC (maintained by IFAE and CIEMAT) in collaboration with ICE-CSIC. It received funding from the Spanish government (grant EQC2021-007479-P funded by MCIN/AEI/10.13039/501100011033), the EU NextGeneration/PRTR (PRTR-C17.I1), and the Generalitat de Catalunya. 

\newline
This work was supported by the French National Research Agency (ANR) under contract ANR-22-CE31-0026.
HZ acknowledgement the supports from the National Natural Science Foundation of China with grant No. 12120101003 and the China Manned Space Project (No. CMS-CSST-2025-A06) and the National Key R\&D Program of China with grant No. 2022YFA1602902.

\newline
We thank the anonymous referee for their helpful comments. 
We thank Yu-Ling Chang and Benjamin Weiner for helpful feedback on the manuscript, and SW thanks Gr\'{e}goire Aufort, Joss Bland-Hawthorn, Cody Carr and Filippo Fraternali for valuable discussions.
SW also thanks Xiu Zhen Chen and Jin Chun Weng, whose teaching of letters and numbers laid the foundation for this work.
\end{acknowledgements}

\section*{Data Availability}
The data used in this analysis will be made public with Data Release 2 (details in \url{https://data.desi.lbl.gov/doc/releases/}). 
Likewise, the catalogue of down-the-barrel absorbers will be available as a DESI Value-added Catalogue after DR2 is released. 
The data corresponding to the figures in this paper are publicly available in a Zenodo repository (\url{https://zenodo.org/records/19674837}). 

\bibliographystyle{aa}
\bibliography{bib}

\appendix
\section{Effect of SNR $>5$ cut on the completeness}
\label{app:snr5}
Before modelling any galaxies, we applied a $\mathrm{SNR}>5$ cut using the rest-frame $\pm 50$\AA\ window around the \ion{Na}{i}~D line (excluding the line itself and the nearby \ion{He}{i} emission feature). 
This threshold is loosely based off previous work \citep{Sato2009}, but in principle, some absorbers may be missed as a result. 
To quantify the impact, we applied our full detection pipeline to $\sim$320,000 spectra with $3 < \mathrm{SNR} < 5$. 
We find that 43 (13) detections exceed our Bayes-factor threshold of $\Delta \ln \mathcal{Z} > 1.0$ ($3.0$) and pass the filtering criteria. 
Scaling these yields an estimated total of $\sim$800 missed cases, corresponding to $\sim$1.7\% of the size of our main sample. 
The estimated $\sim$200 additional detections with $\Delta \ln \mathcal{Z} > 3.0$ represent roughly $\sim$0.8\% of the higher-confidence sample. 
As the $3 < \mathrm{SNR} < 5$ regime contains a large fraction of the overall galaxy population, retaining these lower-quality spectra would substantially increase computational cost while yielding only a marginal increase in genuine detections. 
For this reason, the $\mathrm{SNR}>5$ cut represents a practical balance between completeness and computational feasibility.

\section{Details on the continuum fitting}
\label{app:cont_fit}
The most crucial aspects of fitting down-the-barrel absorption include the accurate estimation of the continuum region around \ion{Na}{i}~D, the modelling of the nearby \ion{He}{i} emission line and most importantly, the separation of interstellar absorption from the stellar photospheric contribution.
Early works relied on polynomials to describe the local continuum around \ion{Na}{i}~D \citep{Rupke2005aSample, Martin2005}, often using the \ion{Na}{i}~D to \ion{Mg}{i}~$b$ equivalent-width ratio to constrain the stellar component.
More recently, stellar population synthesis (SPS) models have been employed to explicitly model the stellar contribution to the \ion{Na}{i}~D profile.
However, these approaches face several challenges in large spectroscopic surveys such as DESI and low-SNR regimes.
At modest SNR, degeneracies between age, metallicity and kinematics limit the robustness of the inferred stellar absorption.
Interstellar absorption itself can also contaminate the stellar spectra used to construct the synthesis libraries, biasing the predicted \ion{Na}{i}~D strength \citep{Machuca2025, Rubin2025}.
In addition, fitting a wide range of stellar templates across large samples is computationally expensive, particularly when repeated for multiple model configurations (e.g. varying [Na/Fe] abundances).
Hence, we choose to fit the systemic component ourselves, as described in \autoref{sec:methods}.

The process of modelling the spectrum introduces systematic uncertainty, which we propagate into the final variance estimate.
As \texttt{FastSpecFit} does not currently return an error estimate, we calculate a fractional error term from the scatter of the observed flux relative to the combined continuum and emission-line model in the wings of the \ion{Na}{i}~D feature, capturing residuals from both the initial stellar continuum and emission-line fits.
The SPS templates adopted by \texttt{FastSpecFit} assume a Chabrier initial mass function \citep{Chabrier2003}, solar metallicity and five age bins spanning 15~Myr to 12.7~Gyr, with constant star formation within each bin.
\texttt{FastSpecFit} is optimised for computational efficiency, which is essential given the size of the DESI data set, but this comes at the cost of a limited range of SPS templates.

To quantify the impact of these assumptions on our continuum estimation, we select 1000 galaxies drawn uniformly from five SNR bins of (5, 7), (7, 10), (10, 15), (15, 25) and ($>25$).
For this subsample, we model the stellar continuum using \texttt{PPXF} as an alternative, applying an identical interpolation procedure across the \ion{Na}{i}~D doublet so that any differences arise from the underlying continuum models and fitted velocity dispersion parameters rather than the treatment of the absorption region itself.
The resulting wavelength-dependent root-mean-square difference between the two algorithms is used to derive an additional fractional variance contribution, which is then applied to the full dataset.

\section{Details on fitting \ion{Na}{i} D}
\label{app:models}
\ion{Na}{i}~D provides a powerful probe of cool neutral gas, but its interpretation is non-trivial. 
In the optically thin limit, atomic physics sets the intrinsic strength ratio of the $D_1$ to $D_2$ lines to 1:2. 
However, as the optical depth increases and the $D_2$ begins to saturate, this ratio increases towards unity. 
In addition, the observed absorption is an average over many sightlines through the galaxy, such that the depth of the lines depends not only on the optical depth of individual absorbers but also on the fraction of the background continuum that they cover. 
As a result, even saturated gas can produce absorption features that do not reach zero intensity if the covering fraction is less than unity. 
These effects introduce degeneracies between optical depth and covering fraction, motivating the use of a partial-covering model to describe the absorption \citep[as in][]{Rupke2005aSample}.

Following the normalisation of the galaxy spectra, any remaining \ion{Na}{i}~D absorption consists of stellar photospheric absorption, interstellar absorption or a combination of both. 
In this framework, we describe six physically distinct cases:
\begin{enumerate}
\item There is no significant absorption, either stellar or interstellar and the normalised continuum is consistent with unity.
\item There is absorption at the systemic redshift arising purely from stellar photospheres.
\item There is absorption at the systemic redshift arising purely from interstellar gas associated with the host galaxy. 
\item There is absorption at the systemic redshift arising from both stars and interstellar gas.
\item There is absorption away from the systemic redshift, indicating inflowing or outflowing gas. 
\item There is absorption both at and away from the systemic redshift, corresponding to a combination of systemic absorption and interstellar absorption from gas flows. 
\end{enumerate}
We note that this list is incomplete, as there can always be additional gas flow components.  

In practice, absorption at the systemic redshift cannot always be unambiguously decomposed into stellar and interstellar components in individual, moderate-SNR spectra. 
We therefore interpret such components phenomenologically, with absorption centred at the systemic velocity representing either stellar absorption, interstellar absorption, or both. 
Naturally, this means we will not be able to recover interstellar absorption if it is near systemic velocity and similar to the stellar velocity dispersion. 
Given that we are primarily interested in outflows and inflows found at negative and positive velocities, respectively, away from the systemic redshift, this limitation does not significantly affect the conclusions of this work.  
The consequences of treating these components identically are discussed in the completeness tests in \autoref{app:PurityCompleteness}. 

Although \ion{Na}{i}~D emission is occasionally observed, incorporating an emission-line component into our models introduces significant complexity. 
Resonant scattering can produce P-Cygni-like profiles in \ion{Na}{i}~D, where blueshifted absorption is accompanied by redshifted emission \citep[e.g.][]{Prochaska2011}. 
However, we find that including an emission component often leads to unphysical results, specifically a degeneracy where excessively strong emission is offset by unphysically deep absorption to mimic the observed profiles. 
These degeneracies are difficult to constrain in individual, moderate-SNR spectra, in contrast to stacking analyses where the average emission component can be more robustly recovered \citep{Concas2019, Roberts-Borsani2019}. 
For these reasons, we currently neglect emission components in our modelling and we show post-hoc in \autoref{sec:results} that this affects only a minority of the sample. 
The fits described hereafter are instead constructed from up to three distinct components, as detailed below.

\paragraph{Null component.}
The null component corresponds to the case with no absorption and is implemented as a degree-0 polynomial. 

\paragraph{\ion{Na}{i}~D systemic absorption.}
We model the systemic absorption using a double-Gaussian model, motivated by the expectation that stellar-dominated \ion{Na}{i}~D absorption produces approximately Gaussian profiles. 
While the absorption from interstellar gas at the systemic redshift may not be perfectly described by a Gaussian profile, we do not aim to characterise the physical properties of this systemic component; this simplified treatment is sufficient for our purposes. 
The continuum-normalised intensity is given by
\begin{equation}
\label{eq:sys_gauss}
\begin{aligned}
I_{\rm sys}(\lambda) &= \left[\,1 
- A_{\mathrm{blue}} \exp\!\left(-\frac{(\lambda - \lambda_{\mathrm{blue}})^2}{2\sigma_{\mathrm{sys}}^2}\right)\right] \\
&\quad \times \left[\,1 
- A_{\mathrm{red}} \exp\!\left(-\frac{(\lambda - \lambda_{\mathrm{red}})^2}{2\sigma_{\mathrm{sys}}^2}\right)\right],
\end{aligned}
\end{equation}
where $A_{\mathrm{blue}}$ and $A_{\mathrm{red}}$ are the amplitudes of the blue and red members of the \ion{Na}{i}~D doublet.
The rest wavelengths of the doublet are fixed to their laboratory values in vacuum and {$\sigma_{\mathrm{sys}}$ is the width of both lines. }
The centroids are constrained by redshift error priors around the systemic velocity of the galaxy. 
This model is intended to capture the combined stellar and interstellar absorption at the systemic redshift, without assigning a unique physical origin to each component. 

\paragraph{\ion{Na}{i} D gas flow absorption.}
In addition to the systemic component, a secondary absorption feature may be observed if gas is flowing either away from (inflow) or towards (outflow) the observer. 
We adopt the partial-covering model of \citet{Rupke2005aSample}, where the continuum-normalised intensity is
\begin{equation}
\label{eq:int_flow_phys}
I_{\rm flow}(\lambda) = 1 - C_f \left[1 - \exp\bigl(-\tau_{\mathrm{blue}}(\lambda) - \tau_{\mathrm{red}}(\lambda)\bigr)\right],
\end{equation}
with $C_f$ representing the covering fraction of the continuum by absorbing gas.
$\tau_{\mathrm{blue}}$ and $\tau_{\mathrm{red}}$ are the optical depths of the blue and red doublet members. 
Each optical depth is assumed to follow a Gaussian profile in the rest frame of the absorbing gas:
\begin{equation}
\label{eq:opt_depth}
\tau(\lambda) = \tau_0 \exp\!\left[-\frac{(\lambda - \lambda_0)^2}{(\lambda_0 b_D / c)^2}\right],
\end{equation}
where $\tau_0$ is the central optical depth, $\lambda_0$ the rest wavelength and $b_D$ the Doppler width. 
The known oscillator strength ratio of 1:2 for the \ion{Na}{i} D doublet is enforced by setting $\tau_{0,\mathrm{red}} = 0.5\ \tau_{0,\mathrm{blue}} $. 
Following \citet{Rupke2005aSample}, we assume $C_f$ is constant with velocity, although in reality the covering fraction may vary within a clumpy medium.

\section{Justification of priors}
\label{app:zerr}
Priors are a foundational part of Bayesian inference: they restrict the parameter space to physically plausible regions and encode external information. 
In this work we adopt predominantly uniform (flat) priors bounded by ranges reported in the literature on down-the-barrel outflows, from dwarf galaxies \citep{SchwartzDwarfOutflow2004} to ULIRGs \citep{Rupke2005aSample, Martin2005}, which together bracket the extremes of observed behaviour. 
These priors are intentionally wide enough that, for parameters with informative data, the posterior is driven by the likelihood, yet constrained enough to exclude unphysical values. 
If stronger prior information is warranted, we adopt informative priors and justify these choices below. 

\subsection{Systemic velocity}
\begin{figure*}
\centering
\includegraphics[width=\linewidth]{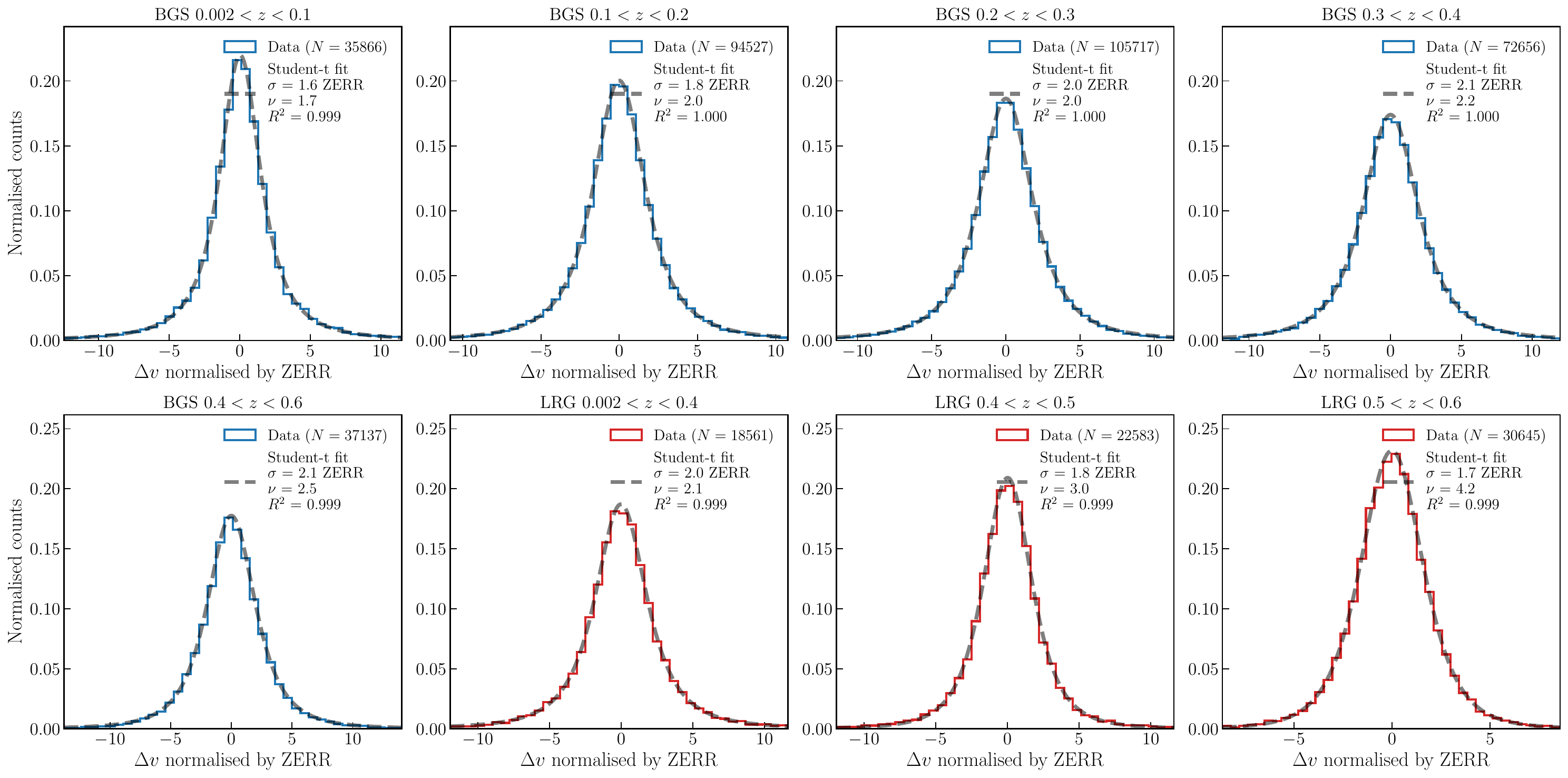}
\caption{
    Velocity offsets normalised by the per-object \zerr value reported by the \texttt{RedRock} pipeline. 
    These velocity differences arise from repeated observations from different nights or tiles of the same target. 
    The breadth of the distribution indicates that \zerr can underestimate the true uncertainty in the systemic redshift.
    }
\label{fig:zErrNorm}
\end{figure*}

While the \textsc{Redrock} pipeline provides a redshift uncertainty ($z_{\rm err}$), a more empirical estimate of the redshift precision can be obtained by measuring the scatter in redshift determinations from repeat observations of the same target taken on different nights or tiles.
Following the approach used during DESI survey validation \citep{Lan2023} and the DESI One-Percent Survey \citep{Yu2024}, we select from DR2 all objects with multiple spectroscopic observations with exposures between 500 and 1,000 seconds (typical DESI exposures). 
The redshift difference is converted to a velocity difference using the first repeated observation.  
The resulting sample comprises roughly $350{,}000$ repeat measurements for BGS galaxies and $70{,}000$ for LRGs, spanning $0.002 < z < 0.6$. 
The resulting error distributions, normalised by the pipeline \zerr values, are shown in \autoref{fig:zErrNorm}. 
Consistent with previous findings, these distributions are not purely Gaussian but exhibit extended tails and are broadly consistent with redshift precisions of $\sim$10~\kms for BGS galaxies and $10$--$60$~\kms for LRGs \citep{Lan2023,Yu2024}. 
We therefore model the empirical distributions using Student's $t$ functions after excluding catastrophic errors with $|\Delta v| > 1000$~\kms which constitute $<1\%$ of repeat measurements in most bins. 
The Student's $t$ fit characterises both the width ($\sigma$) and the non-Gaussianity via the degrees of freedom parameter $\nu$ (with $\nu=1$ reducing to a Cauchy profile and $\nu\to\infty$ converging to a Gaussian).

For each galaxy, we construct a prior on the systemic velocity using its pipeline redshift, redshift error and galaxy type.
Allowing the systemic velocity to vary within this data-driven prior, rather than fixing it, reduces false positives that would otherwise arise from small redshift offsets being misinterpreted as real gas motions.
We note that ongoing analysis (He et al. in preparation) indicates that estimating uncertainties from repeat-observation redshift differences overestimates the true intrinsic redshift error distribution.
This occurs because such estimates implicitly assume that one of the repeat measurements represents the true redshift, whereas in reality both measurements are noisy realisations drawn from a distribution centred on the true value.
Therefore, the priors adopted in this work are conservative, in the sense that they allow for a broader range of systemic velocities than is likely present in the data.

\paragraph{Systemic line ratio} 
{We constrain the \ion{Na}{i} D doublet ratio ($R_{12} =$ D$_1/$D$_2$) of the systemic component to the range $0.75$--$1.0$. }
This range is physically motivated by the expectation that absorption arising from the stellar photosphere and the systemic interstellar medium is optically thick \citep{Heckman2000}. 
{In the stellar atmospheres of cool stars, $R_{12} =$ 0.8--1.0  \citep{Heckman2000}.}
Similarly, the systemic ISM gas in galactic discs is also saturated. 
Treating these two sources as a single component is further supported by the fact that neutral interstellar gas and stars are generally co-spatial and share comparable scale heights in galactic discs \citep[e.g.][]{Leroy2008, Bagetakos2011}. 
By adopting this prior, we model the combined systemic contribution as a single feature \citep[see][]{Rubin2014}, preventing the fit from diverging toward the unphysical 1:2 ratio at the optically thin limit.

\paragraph{Systemic velocity dispersion}
The velocity dispersion of the systemic component is constrained via a two-tiered approach. 
For approximately 70\% of our sample, we adopt the stellar kinematic measurements provided by \texttt{FastSpecFit}. 
As \texttt{FastSpecFit} uses the full spectral continuum and multiple stellar absorption features, it provides a more robust estimate of the stellar velocity dispersion than stand-alone fitting of the \ion{Na}{i} D doublet. 
In these cases, we apply a uniform prior centred on $\sigma_{*, \rm FSF}$ with a half-width of $3\epsilon_{*, \rm FSF}$, where $\epsilon_{*, \rm FSF}$ represents the $1\sigma$ measurement uncertainty. 
These priors are designed to be sufficiently broad to accommodate a systemic interstellar component that may exhibit a slightly different dispersion than the stellar population. 
A post-hoc comparison confirms that the values recovered from our nested sampling are in excellent agreement with the \texttt{FastSpecFit} estimates, exhibiting a $1\sigma$ dispersion of $19$ \kms.
For the remaining 30\% of sources where such measurements are unavailable, we use a conservative uniform prior ranging from 30 to 450 \kms, encompassing the expected velocity dispersions for systems ranging from low-mass dwarfs to massive early-type galaxies.

\paragraph{Gas flow velocity:}
The velocity of the gas flow is restricted to the range $[-1100, 1100]$ \kms.
Outflows have been observed down the barrel at extreme velocities of $>1000$ \kms at $z \sim 0.6$ \citep{Tremonti2007}.
Inflows are less well studied, but available measurements suggest generally lower speeds of $\lesssim 200$~\kms \citep{Rubin2012, Roy2021, Moghni2026}, with occasional cases reaching $\sim$500~\kms \citep{Martin2012}.
Given the limited constraints on inflow velocities, we adopt a broad prior that covers the full plausible range of both inflow and outflow speeds, which is implemented as a symmetric and uniform interval about zero.

This prior choice also explains why the \texttt{flow} model is not used for candidate selection.
When compared with the systemic model, the flow model is almost always favoured once the fit moves even slightly away from the systemic redshift.
The Student-$t$ prior of the systemic model concentrates nearly all of its probability density very close to zero velocity, so its prior support drops rapidly at larger offsets.
In contrast, the \texttt{flow} model adopts a broad uniform prior, which maintains support across the full velocity range.
As a result, small changes in the velocity bounds of the \texttt{flow} prior can lead to appreciable changes in the evidence ratio between models, making the comparison sensitive to arbitrary prior choices rather than the data alone.
While a more informative prior on gas-flow velocities could alleviate this issue, we currently lack a well-motivated prescription.
We therefore adopt a uniform prior and exclude the \texttt{flow} model from the candidate-selection step.

\paragraph{Gas flow dispersion:}
Similar to the flow velocity, we use the extreme values in the literature to set the boundaries of our uniform prior \citep{Krug2010, Rubin2014}. 
Hence, we allow Doppler parameters between 10 and 500~\kms, wide enough to encompass both narrow components and the broadest features reported in down-the-barrel studies.

\section{Nested sampling for candidate selection}
\label{app:nest_cand}
The most time-consuming stage of our analysis is the identification of candidate down-the-barrel absorbers.
Given the large parent sample of approximately six million spectra with SNR $>5$, we adopt nested sampling for candidate selection.
While this approach is more computationally expensive than simpler fitting methods, it provides a more reliable basis for model comparison and robust discrimination between genuine absorption features and noise-driven solutions.

The computational cost of nested sampling scales with the number of live points, $n_{\rm live}$, such that this parameter controls the balance between run time and the accuracy of the evidence estimate.
In principle, $n_{\rm live}$ should increase with the number of free parameters, $n_{\rm dim}$, as the explored prior volume grows with dimensionality.
For the purposes of an initial candidate search across a very large sample, we adopt a conservative scaling of $n_{\rm live} = 20 \times n_{\rm dim}$, which significantly reduces CPU cost while remaining adequate for robust model selection.

\begin{figure}
\centering
\includegraphics[width=\linewidth]{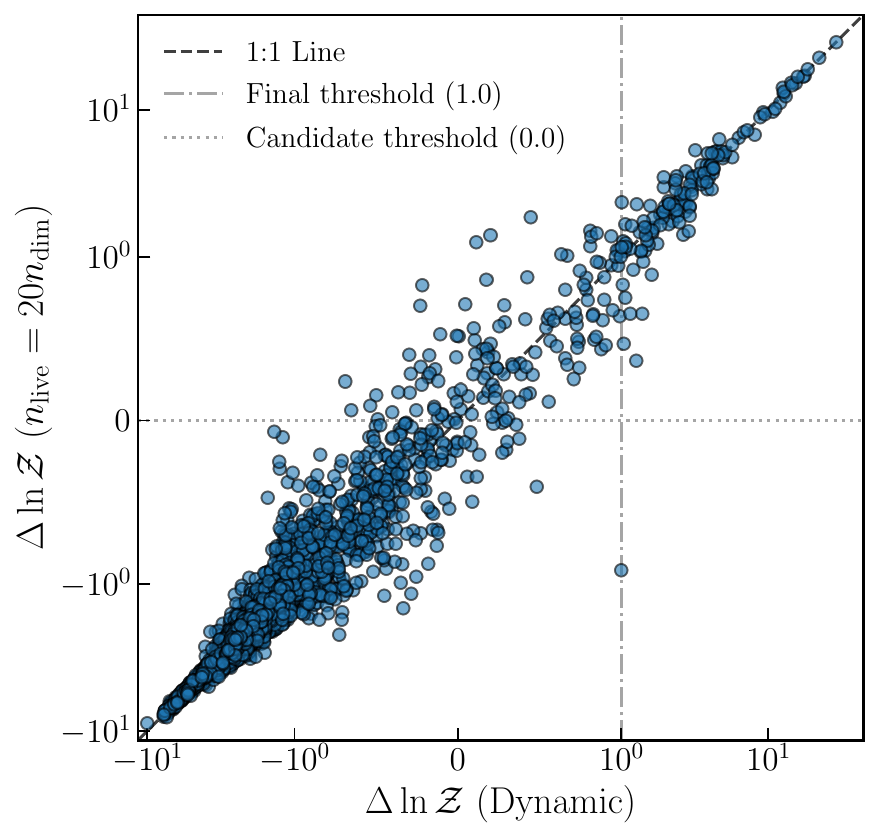}
\caption{
Comparison of $\Delta\ln\mathcal{Z}$ values measured using the candidate-selection configuration ($n_{\rm live}=20\,n_{\rm dim}$) and a higher-accuracy dynamic nested sampling run. 
The dashed line shows the one-to-one relation. 
The close agreement, with only modest scatter, demonstrates that the reduced $n_{\rm live}$ configuration provides sufficiently accurate evidence estimates for robust model selection. 
The vertical line indicates the final detection threshold used while the dotted horizontal line shows the candidate-selection threshold. 
}
\label{fig:evidence_compare}
\end{figure}

In \autoref{fig:evidence_compare}, we compare evidence differences, expressed as $\Delta\ln\mathcal{Z}$ between the best-fitting outflow model and baseline model, obtained with our candidate-selection configuration ($n_{\rm live}=20\,n_{\rm dim}$, \texttt{dlogz} $=0.1$) and with a higher-accuracy dynamic nested sampling run used as a reference ($n_{\rm live}=500$, \texttt{batch}$=100$, \texttt{dlogz\_init} $=0.05$, \texttt{pfrac} $=0.1$).
We find that the $\Delta\ln\mathcal{Z}$ values are consistent between the static and dynamic nested sampling runs, closely following the one-to-one relation with only modest scatter, indicating that our choice of $n_{\rm live}$ is sufficient for reliable model comparison in the candidate selection stage.

\begin{figure}
\centering
\includegraphics[width=\linewidth]{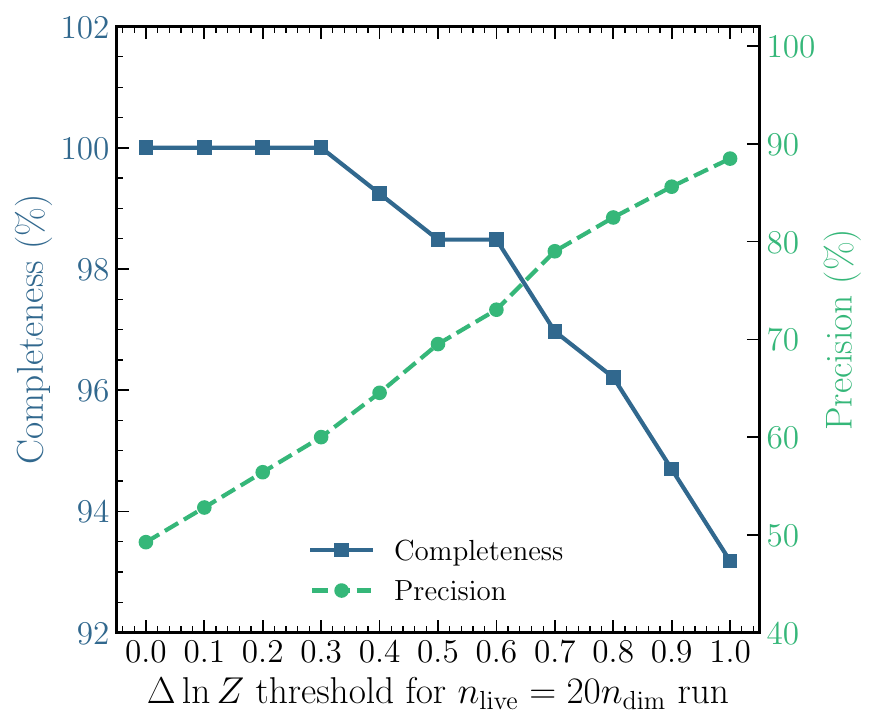}
\caption{
    Calibration of the evidence threshold used for candidate selection, based on a test sample of $\sim$20{,}000 spectra.
    The completeness (solid line) is defined as the fraction of $\Delta\ln\mathcal{Z}>1.0$ detections in the high-accuracy dynamic nested sampling run that are recovered by the lower-cost static run, as a function of the static-run evidence threshold.
    The precision (dashed line) denotes the fraction of static-run detections that are also identified in the dynamic run.
}
\label{fig:nlivecomp}
\end{figure}

Figure~\ref{fig:nlivecomp} shows the fraction of sources with $\Delta\ln\mathcal{Z}>1.0$ in the dynamic run that are recovered in the static run, as a function of the evidence threshold applied to the latter.
Using identical thresholds yields a completeness of $\sim$90\%, but for candidate selection we prioritise completeness over precision.
We therefore adopt a permissive threshold of $\Delta\ln\mathcal{Z}>0.0$.
Although completeness reaches 100\% by $\Delta\ln\mathcal{Z}\simeq0.3$ in this test sample, we expect increased statistical variance when scaling to the full dataset.
The resulting decrease in precision increases downstream computational cost, but this remains negligible compared to the cost of performing nested sampling on the full parent sample.

\section{Completeness and Purity}
\label{app:PurityCompleteness}

\subsection{Injection and recovery tests}
\label{app:inject}

We assess completeness by injecting synthetic \ion{Na}{i}~D absorption profiles into real DESI spectra. 
Target spectra are chosen to be free of \ion{Na}{i}~D absorption by requiring $\Delta \ln \mathcal{Z} > 3$ for the \texttt{null} model relative to all other models and are then inspected visually.  
For each of four continuum SNR bins around the injection wavelength, (5, 7), (7, 10), (10, 15) and (15, 25), we draw 200 spectra, giving a broad sampling of instrumental and observational conditions.

Absorption models of types \texttt{sys} and \texttt{sys\_flow} are generated using the parameter ranges listed in the final column of \autoref{tab:priors} and each model is convolved with the DESI resolution matrix prior to injection.  
The velocity offset of the stellar component from the \textsc{RedRock} redshift is drawn from the empirical redshift-error distribution that we use to construct our priors.  
Profiles are injected multiplicatively into the spectra and we discard injections with total equivalent widths $<0.25$~\AA\ or $>6.0$~\AA.  
After these cuts, the procedure produces roughly $1.2$ million synthetic profiles.

To explore noise effects, we adopt a simple two-regime model based on the Legacy Survey $r$-band magnitude: objects brighter than $r = 19$~mag are treated as photon dominated, while fainter objects are treated as sky dominated.  

Injected spectra are processed through the same pipeline as the real data.  
We use static nested sampling for computational efficiency, noting that it agrees well with dynamic sampling (\autoref{fig:nlivecomp}).  
We inject only single-component models and assume no pre-existing absorption.  
Using real spectra ensures that sky residuals, tellurics and cosmic rays are naturally included.

\subsection{Completeness}
We quantify completeness by recovering injected profiles using the same criteria applied to the real data.  
A detection requires $\Delta \ln \mathcal{Z} > 1.0$ for the \texttt{sys\_flow} model relative to both \texttt{null} and \texttt{sys}, and recovery of the injected velocity within 50--100 \kms\ depending on $|v_{\rm flow}|$.

A key caveat is that completeness reflects only the sampled parameter grid.  
It does not represent the true astrophysical population, which is unknown, and does not fully capture redshift-dependent effects such as tellurics or sky emission.  
It should therefore be interpreted as a measure of pipeline sensitivity.

\begin{figure*}
    \centering
    \includegraphics[width=1.0\linewidth]{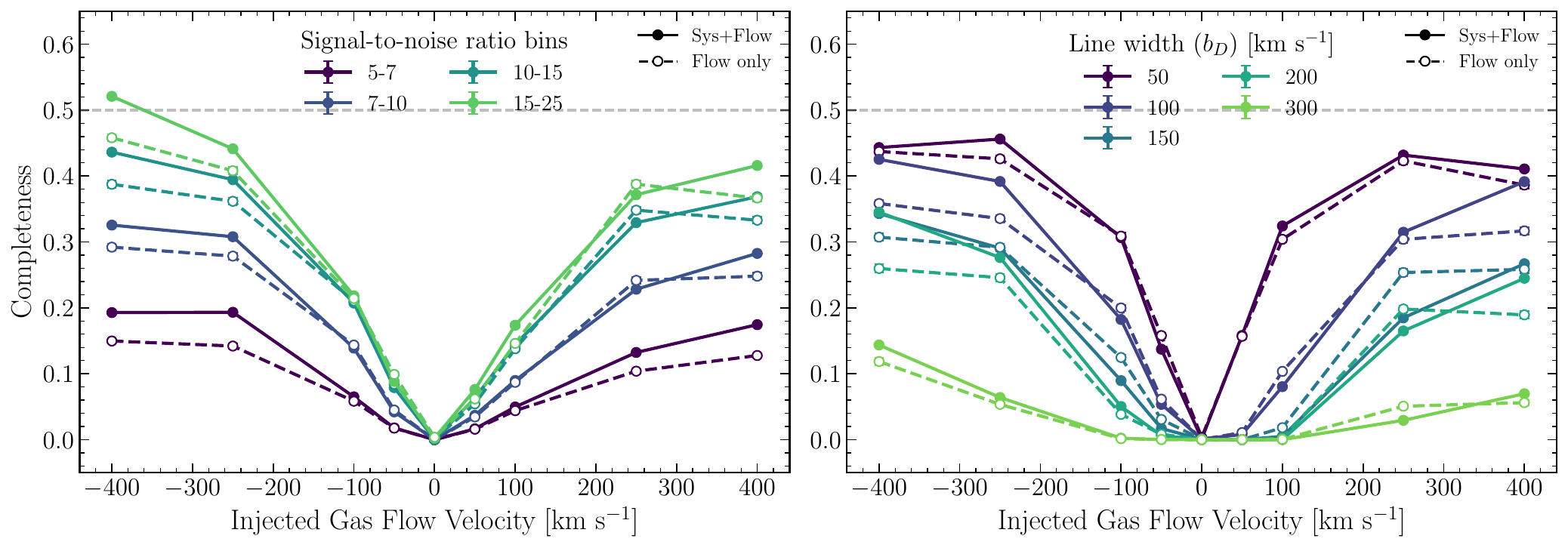}
    \caption{
    Completeness as a function of injected gas-flow velocity for different continuum SNR bins (left) and line widths $b_D$ (right).  
    Solid lines show injections with a systemic component and dashed lines pure flow injections.
    }
    \label{fig:completeness_1D}
\end{figure*}

In \autoref{fig:completeness_1D}, completeness increases with SNR and is higher for narrower components.  
Broad features are more difficult to recover due to blending and the imposed depth threshold.  
Completeness is also reduced near systemic velocity, particularly for low SNR and large $b_D$, where blending with the systemic component becomes significant.

We find an asymmetry between inflows and outflows: redshifted components are less frequently recovered than blueshifted ones.  
{This is caused primarily by blending within the \ion{Na}{i}~D doublet itself, rather than blending with another species.  
Since the two doublet lines are separated by only $\sim$300 \kms, redshifted flow components move the stronger \ion{Na}{i}~D2 line into the partially blended systemic D2 and D1 region, reducing the contrast of the inflow signature.  
Blueshifted components instead move the stronger D2 line away from this blended region, making them easier to identify at fixed SNR, $b_D$, optical depth and covering fraction \citep{Rupke2005bAnalysis, Martin2005}.}

{
This asymmetry implies that some inflows are likely missed.  
However, estimating the total number of missed inflows is model dependent.  
Such an estimate would require knowledge of the intrinsic distributions of inflow velocities, velocity widths, optical depths and covering fractions.  
It would also require assumptions about whether inflow-hosting galaxies are representative of the parent sample, which is unlikely given that our detections suggest that inflows preferentially occur in more massive galaxies.  
Constraining a completeness-corrected inflow-hosting fraction, and hence the global covering fraction of inflows, is the focus of ongoing work.}

Completeness is also influenced by model assumptions, with slightly higher recovery for \texttt{sys\_flow} than pure \texttt{flow} cases due to prior-volume effects.

\subsubsection{Comparison with the literature}
\label{app:comparelit}
In addition to validating our methodology using synthetic spectra, we also compare our measurements to previously studied galaxies that appear in the DESI footprint.
We cross-match our SNR $> 5$ sample with the datasets of \citet{Martin2005} and \citet{Rupke2005aSample}, identifying 31 galaxies in common. 
Although this overlap is naturally biased toward infrared-luminous, intensely star-forming systems that may possess stronger and more readily detectable outflows, it provides a useful benchmark for assessing our ability to recover the most extreme cases reported in the literature.

For these 31 galaxies, we find good agreement between our DESI-derived flow measurements and those reported in previous work for 22 objects (71\%).  
Among the galaxies with outflows, we recover 11/20 (55\%), while the remaining 11 consistent cases do not show measurable flows in both data sets.  
This level of agreement is broadly consistent with expectations based on the difference in data quality: the median SNR per \AA\ of our DESI spectra is 11.7, compared to 26 in \citet{Rupke2005aSample}.  
From our completeness tests, such a decrease in SNR corresponds to an expected $\sim$40\% drop in completeness in the $v_{\rm flow} \leq 250$~\kms regime, roughly matching the fraction of cases in which we do not recover the published outflows.  
In several sources we find marginal evidence for an additional component ($0 < \Delta \ln \mathcal{Z} < 1$); these fall below our adopted detection threshold but would likely be recovered with higher-SNR data.  
Some mismatches also arise from methodological differences between our Bayesian framework and traditional frequentist analyses, particularly in the treatment of systemic redshift uncertainties.

\subsection{Purity}
We distinguish between two complementary notions of purity. 
The first concerns the absence of direct false positives, in which an apparent \ion{Na}{i}~D absorption feature is not astrophysical in origin (e.g.\ residual sky lines, reduction artefacts, or continuum-fitting failures).  
This form of purity is addressed by the candidate-filtering procedure described at the end of \autoref{sec:methods} and can be assessed reliably by visual inspection.  
Among 1000 visually vetted detections, fewer than 3\% are clear false positives of this type.

The second notion of purity is statistical and relates to model selection rather than astrophysical reality.  
In our framework, the strength of an absorption component is effectively a continuous quantity and the identification of a gas flow requires the adoption of a threshold at which a component is deemed to be significantly detected.  
We implement this through a Bayes factor criterion, such that a flow component is included only when the evidence for a more complex model exceeds that of a simpler nested alternative by $\Delta \ln\mathcal{Z} > 1$.  
This corresponds to the data being approximately three times more probable under the model including the additional component.

It is important to note that any alternative procedure, including visual inspection, would implicitly impose a similar threshold on the strength or significance of the absorption.  
Even a perfectly unbiased human classifier would primarily act to shift this effective detection threshold, rather than to fundamentally alter the classification of strong detections.  
As such, the statistical purity of the sample is intrinsically tied to the choice of this threshold, and cannot be defined independently of it.

For this reason, we do not attempt to assign a single numerical value to the statistical purity of the sample. 
Instead, we adopt a conservative evidence threshold designed to favour robustness over completeness, thereby minimising the inclusion of marginal components.  
Crucially, our analysis focuses on population-level trends and relative comparisons across galaxy properties, rather than on the interpretation of individual components.
Such trends are insensitive to a small fraction of borderline detections.
We therefore conclude that, while the exact statistical purity cannot be uniquely quantified, the adopted methodology yields a sample that is sufficiently robust for the purposes of this study.

\section{Degeneracy between covering fraction and optical depth}
\label{app:tau_degen}
\begin{figure}
\centering
\includegraphics[width=\linewidth]{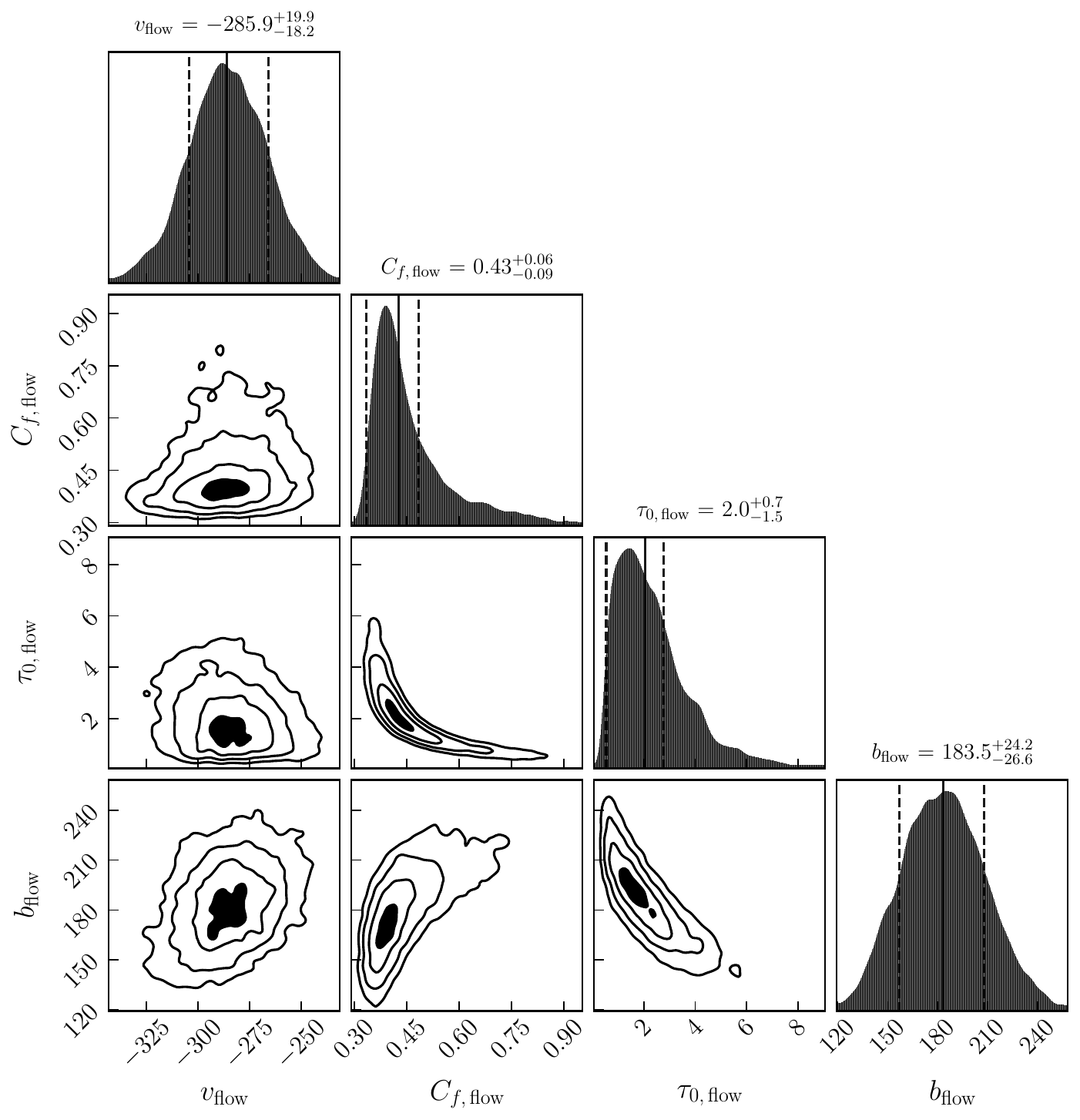}
\caption{
Corner plot showing the posterior distributions for a representative galaxy spectrum best fitted by the \texttt{flow} model (the zero-order normalisation parameter $K$ is omitted for clarity).
The joint posterior for the covering fraction ($C_{\rm f,flow}$) and central optical depth ($\tau_{0,{\rm flow}}$) displays the characteristic curved, `banana-shaped' degeneracy commonly encountered in absorption-line modelling.
One-dimensional marginal distributions along the diagonal show the median and 68-percentile intervals for the inferred parameters.
}
\label{fig:corner}
\end{figure}

A pronounced degeneracy is evident between the covering fraction $C_{\rm f}$ and the line-centre optical depth $\tau_0$.
This degeneracy arises because a wide range of $(C_{\rm f}, \tau_0)$ pairs can produce absorption profiles of similar depth.
In partial-covering models, increases in $\tau_0$ can be compensated by reductions in $C_{f}$ and vice versa.
The likelihood is therefore elongated along this direction in parameter space, producing the curved joint posterior seen in \autoref{fig:corner}.

In principle, this degeneracy can be mitigated by using information from doublets or multiple transitions of the same ion.
In practice, however, the small wavelength separation of the \ion{Na}{i} D doublet limits the extent to which the two components can be disentangled, especially at moderate resolution, making this approach less effective than for more widely separated doublets such as \ion{Mg}{ii}.

\end{document}